\documentclass[aip,pop,preprint,groupedaddress,preprintnumbers,showpacs,amsmath,amssymb]{revtex4-1}

\usepackage{graphicx}
\usepackage{dcolumn}
\usepackage{bm}

\usepackage{units}

\usepackage{color}

\begin{document}


\title{Enhanced Stopping of Macro-Particles in Particle-in-Cell Simulations}


\author{J. May and J. Tonge}
\affiliation{Department of Physics \& Astronomy, University of California, Los Angeles, CA 90095}
\author{W. B. Mori}
\affiliation{Department of Electrical Engineering, University of California, Los Angeles}
\affiliation{Department of Physics \& Astronomy, University of California, Los Angeles, CA 90095}
\author{F. Fiuza}
\affiliation{Lawrence Livermore National Laboratory, CA 94550}
\author{R. A. Fonseca}
\author{L. O. Silva}
\affiliation{GoLP/Instituto de Plasma e Fus\~{a}o Nuclear, 1049-001 Lisboa, Portugal}
\author{C. Ren}
\affiliation{University of Rochester}

\date{\today}

\begin{abstract}
We derive an equation for energy transfer from relativistic charged particles to a cold background plasma appropriate for finite-size particles that are used in particle-in-cell simulation codes.
Expressions for one-, two-, and three-dimensional particles are presented, with special attention given to the two-dimensional case.
This energy transfer is due to the electric field of the wake set up in the background plasma by the relativistic particle.
The enhanced stopping is dependent on the $q^2/m$, where $q$ is the charge and $m$ is the mass of the relativistic particle, and therefore simulation macro-particles with large charge but identical $q/m$ will stop more rapidly.
The stopping power also depends on the effective particle shape of the macro-particle.
These conclusions are verified in particle-in-cell simulations.
We present 2D simulations of test particles, relaxation of high-energy tails, and integrated fast ignition simulations showing that the enhanced drag on macro-particles may adversely affect the results of these simulations in a wide range of high-energy density plasma scenarios.
We also describe a particle splitting algorithm which can potentially overcome this problem and show its effect in controlling the stopping of macro-particles.
\end{abstract}


\maketitle

\section{\label{sec1} Introduction}
Particle-In-Cell (PIC) simulations \cite{dawson:1983,birdsall} are a useful tool for studying nonlinear and kinetic physics such as those found in laser-solid interactions \cite{silva:2004,fiuza:2012}, fast ignition inertial fusion energy \cite{tabak:1994,chrisman:2008,tonge:2009,kemp:2010,may:2011}, and relativistic collisionless shocks \cite{spitkovsky:2008,martins:2009}.
Since PIC codes are based on first principles they are a preferred tool to explore physics in regimes where details of the distribution function affect the overall behavior of the system, where there have not been many experiments, or where experimental results are not well understood.
For instance, in laser-solid plasma interactions, fast ignition, and relativistic shock studies fully kinetic simulations that resolve the electron dynamics are necessary.
In these problems the flow of relativistic electrons in a background plasma is an important process.
However, the physical scale length that needs to be resolved in these systems is the collisionless skin depth and not the Debye length.
As the Debye length needs to be resolved in order to avoid numerical grid heating which results from aliasing, the resolution used in such simulations is sometimes much finer or the electron temperature is much higher than necessary.
The use of higher order splines, and current smoothing and compensation can eliminate grid heating permitting larger cells and/or lower temperatures to be used \cite{birdsall,sentoku:2008}.
The use of larger cells also permits larger simulations that model the full spatial domains of the problems of interest.
In order to carry out such large simulations in multiple dimensions it is common to limit the number of particles per cell, which is generally chosen at a value that keeps spurious finite size particle collisions and noise to a sufficiently small value.

Here we describe and quantify another effect that needs to be considered in PIC simulations when relativistic electrons are present \cite{tonge:2010,kemp:2013x}.
This work was originally motivated by studies of fast ignition relevant plasmas \cite{tonge:2009}.
In these simulations a high intensity laser self-consistently generates a distribution of relativistic electrons with densities up to $\sim 10^{22}$cm$^{-3}$ moving through a $10^{23}$cm$^{-3}-10^{24}$cm$^{-3}$ density background plasma \cite{chrisman:2008,tonge:2009,kemp:2009}.
In a previous paper \cite{tonge:2009} we observed an anomalous relaxation of the tail in the distribution function of these high-energy electrons in $10^{23}$cm$^{-3}$ density plasma;
we ascribed this relaxation to the effect of plasmon emission, or wakes, in a turbulent plasma.
Although plasmon emission is a physical effect, in this paper we show that its importance on a single particle is enhanced by the use of macro-particles in PIC codes.
We show that the stopping power of relativistic electrons due to plasmon emission scales as $q^2/m$, where $q$ is the charge and $m$ is the mass of the simulated relativistic electrons.
In PIC simulations the ratio $q/m$ of the simulated particles is kept consistent with real electrons, but the mass and hence the charge can be many times greater, thus strongly affecting the energy exchange between relativistic electrons and background plasma.
In addition, the stopping power will depend on the dimensionality of the simulation and on the shape of the macro-particles.
This artificially high plasmon emission can be controlled by using more particles per cell, larger cells, and higher order particle shapes.
This is similar to how collisions and fluctuations are controlled in a PIC code.
In this paper, we derive expressions for the stopping power of relativistic electrons in PIC simulations, and discuss how parameters can be chosen to reduce this effect with an emphasis on two-dimensional (2D) simulations of fast ignition relevant plasmas.
These results are also relevant to general laser-solid interaction and relativistic shock simulations.

This paper is organized as follows.
In Section \ref{sec2} the equations for the macro-particle stopping of high energy particles in 2D PIC simulations by energy transfer through wakes are derived.
The 1D and 3D cases are also discussed.
In Section \ref{sec3} the importance of macro-particle stopping in fast ignition simulations is discussed and the stopping of single particles in a nominal fast ignition plasma is analyzed.
Section \ref{sec4} addresses the relaxation of high-energy tails of the electron distribution function due to macro-particle stopping.
In Section \ref{sec5} we turn to full-scale fast ignition simulations and analyze how varying the number of particles per skin depth in isolated target simulations changes the results.
Section \ref{sec6} outlines a particle splitting technique we have developed to reduce the effect of stopping by reducing the size of high energy electrons.
Finally, in Section \ref{sec7} we summarize our results.

\section{\label{sec2} Theory}
A charged particle moving faster than the thermal velocity in a plasma forms density wakes.
The electric field of the wake at the location of the particle will do work on the particle and slow it down.
To find the energy loss due to these wakes,
we will consider a highly relativistic point electron moving through a cold fluid background plasma,
and will solve for the electric field of the wake, following the methods used in the plasma-based accelerator community \cite{katsouleas:1987,mori:1988}.
We present results for the wake and electric field on the particle for 1D, 2D, and 3D.
We then consider the effect of finite sized particles on the stopping,
with an emphasis on 2D as it is presently the most common case. 

We start by considering relativistic particles with charge $\rho_b$ moving near the speed of light ($c$) in a cold fluid plasma, following the approach in Ref. \cite{katsouleas:1987}.
The linearized fluid equations for the background plasma are:
\begin{equation} 
\frac{\partial \vec{v_1}}{\partial t}=-\frac{e}{m_e} \vec{E},
\label{eq:force}
\end{equation}
\begin{equation} 
\frac{\partial n_1}{\partial t}=-n_0 \vec{\nabla} \cdot \vec{v_1},
\label{eq:cont}
\end{equation}
\begin{equation} 
\vec{\nabla} \cdot \vec{E}=-4\pi e {n_1} + 4\pi q n_b,
\label{eq:Gauss}
\end{equation}
where $e$ is the charge of an electron, $m_e$ is its mass, $v$ is the electron fluid velocity, $E$ is the electric field, $n$ is the plasma density, the subscripts 0 and 1 indicate zeroth and first order quantities respectively, and the subscript b corresponds to quantities for the relativistic particle species.
We next consider the response for a single beam particle moving in the z direction with a speed $v_b$, in which case 
\begin{equation} 
 q n_b =q \delta(\vec{r}-\vec{v_b}t)= q\delta(z-v_bt)\delta(\vec{r}_{\perp})
\label{eq:rhob}
\end{equation}
where $\delta$() is the Dirac delta function, $q$ is the magnitude of the charge on the test particle, $\vec{r}$ the position vector, and $\vec{r}_{\perp}$ is the part of the position vector perpendicular to the direction of motion.
The response to this point particle can also be viewed as the Green's function response of the electric field when one is calculating the response to a continuous beam distribution. 
We will also use Faraday's law 
\begin{equation} 
-\vec{\nabla} \times \vec{E}=\frac{1}{c}\frac{\partial \vec{B}}{\partial t},
\label{eq:faraday}
\end{equation}
and Ampere's law where we have substituted for the current
\begin{equation} 
\vec{\nabla} \times \vec{B}=-\frac{4\pi e n_0}{c} \vec{v_1} + \frac{4\pi q n_b}{c} \vec{v_b} + \frac{1}{c} \frac{\partial \vec{E}}{\partial t}.
\label{eq:ampere}
\end{equation}

We combine these equations to get
\begin{equation} 
\frac{\partial^2 n_1}{\partial t^2 } + \omega_p^2 n_1 =\omega_p^2 \frac{q}{e} n_b =\frac{\omega_p^2}{v_b}\frac{q}{e}\delta(\vec{r}_{\perp})\delta(t-\frac{z}{v_b}).
\end{equation}
where $\omega_p = \sqrt{4 \pi n_0 e^2/m_e}$ is the unperturbed plasma frequency.

Due to causality the density perturbation in front of the relativistically moving test particle must vanish, so the density response is 
\begin{equation} 
n_1 = \frac{\omega_p q}{v_b e}\delta(\vec{r}_{\perp})\eta(t-\frac{z}{v_b})\mathrm{sin}(\omega_p(t-\frac{z}{v_b}))
\end{equation}
where $\eta()$ is the Heaviside step function.
To find the electric field, we use  equations \eqref{eq:force},\eqref{eq:faraday}, and \eqref{eq:ampere} to derive a wave equation for the electric field, 
\begin{equation} 
 -\frac{1}{c^2}\frac{\partial^2  \vec{E}}{\partial t^2 }+\nabla^2 \vec{E}-\vec{\nabla}\vec{\nabla}\cdot \vec{E} =  k_p^2 \vec{E} +\frac{4\pi}{c^2} \frac{\partial \vec{v_b}q n_b}{\partial t}.
\end{equation}
where $k_p = \frac{\omega_p}{c}$ is the plasma wave number.

Next, we substitute from Gauss's law \eqref{eq:Gauss} and assume the test particle's velocity remains very close to $c$, i.e. $\vec{v_b} = c \hat{z}$, where $\hat{z}$ is the unit vector in the z direction.
We concentrate on the component of the electric field in the $\hat{z}$ direction, and use the fact that both $E_z$ and $\rho_b$ are therefore functions of ($z-ct$) such that $\frac{\partial}{\partial t} + c\frac{\partial}{\partial z} = 0$, to obtain

\begin{equation} 
(\nabla^2_{\perp}-k_p^2)E_z= -4\pi k_p q \frac{\partial}{\partial z}[\delta(\vec{r_{\perp}})\eta(t-\frac{z}{c})\mathrm{sin}(\omega_p(t-\frac{z}{c}))]
\end{equation}

The solution for $E_z$ can then be written as
\begin{equation} 
E_z= G_R(\vec{r_{\perp}}) 4\pi k_p^2q \eta(t-\frac{z}{c})\mathrm{cos}(\omega_p(t-\frac{z}{c}))
\label{eq:E_z_anyD}
\end{equation}
where
\begin{equation} 
(\nabla^2_{\perp}-k_p^2)G_R(\vec{r_{\perp}})= \delta(\vec{r_{\perp}}).
\label{eq:Greens}
\end{equation}

The solutions to Eq. \eqref{eq:Greens} depend on the dimensionality of the problem.
In 1D, the $\delta(r_\perp)$ in Eq. \eqref{eq:rhob} should be replace by 1, $\nabla^2_{\perp}$ vanishes,  and $q$ is in units of charge per unit area, so that $G_{R\,1D} = -1/k_p^2$.
This yields an expression for the electric field
\begin{equation}
E_{z\,1D}(z)  = -4 \pi \bar{\bar{q}} \eta(t - \frac{z}{c}) \mathrm{cos}( \omega_p ( t - \frac{z}{c} ) ) 
\label{eq:Ez1D}
\end{equation}
where $\bar{\bar{q}}$ is the change per unit area. 

In 3D, the solution to Eq. \eqref{eq:Greens} 
can be shown to be the Modified Bessel Function of the Second Kind \cite{abramowitz:1970}
\begin{equation}
G_{R\,3D}(\vec{r_{\perp}}) = -\frac{1}{2\pi} K_0(k_p |r_{\perp}|)
\end{equation}
which yields the Green's function for the electric field behind a test charge
\begin{equation} 
E_{z\,3D}(r_{\perp},z)=  -2q k_p^2 K_0(k_p r_{\perp}) \eta(t-\frac{z}{c})\mathrm{cos}(\omega_p(t-\frac{z}{c})),
\label{eq:Ez}
\end{equation}
The stopping power on a charge is the Green's function evaluated at the particle.
For point particles the value of the step function is \nicefrac{1}{2} at the position of the particle.
From Eq. \eqref{eq:Ez}, in 3D the electric field diverges logarithmically at the origin, which violates the linearity condition.
The stopping power would also diverge logarithmically.
Therefore, if this response is viewed as the electric field from a real point electron then the result cannot be valid near the origin.
However, when viewed from the stand point of a finite size particle (or beam), what matters is the solution for $E_z$ obtained by integrating the Green's function  over the particle's shape.
As long as this electric field is finite then the use of linear theory can still be valid.
We next carry out this integral in detail for the 2D case and comment on the 1D and 3D cases at the end of the section.

In 2D (slab) geometry, where $\nabla^2_{\perp} \rightarrow \frac{d^2}{d r_\perp^2}$ and $\delta(\vec{r_\perp}) = \delta(r_\perp)$ it is readily verified \cite{mori:1988} that the solution to Eq. (\ref{eq:Greens}) is given by setting $G_{R\,2D}(r_{\perp})=-\frac{1}{2k_p}e^{k_pr_{\perp}}$ for $r_{\perp}<0$ and  $G_{R\,2D}(r_{\perp})=-\frac{1}{2k_p}e^{-k_p r_{\perp}}$ for $r_{\perp}>0$.
Thus the electric field is
\begin{equation} 
E_{z\,2D}(r_{\perp},z)=  -2\pi e^{-k_p\mid r_{\perp}\mid} k_p\bar{q} \eta(t-\frac{z}{c})\mathrm{cos}(\omega_p(t-\frac{z}{c})),
\label{eq:Ez2d}
\end{equation}
where $\bar{q}$ is the charge per unit length.
Unlike the 3D case, this electric field remains finite at the origin allowing direct analysis without specifying a particle shape.
Using \nicefrac{1}{2} for the value of $\eta (0)$, the energy loss for a point (line of charge in 2D) particle is given by
\begin{equation} 
\frac{d\epsilon_{2D}}{dt}=-\pi  \omega_p \bar{q}^2
 \label{eq:eloss}
\end{equation}
where $\epsilon$ is the energy of the test particle.
Dividing by $\omega_p \bar{m}c^2$ we get 
\begin{equation} 
\frac{d\gamma_{2D}}{d\omega_pt}=-\pi\frac{\bar{q}^2}{\bar{m}c^2}.
 \label{eq:gloss}
\end{equation}
where  $\gamma$ is the relativistic Lorentz factor of the test charge.
Equation \eqref{eq:gloss} shows that the stopping power (distance) for a relativistic particle in a cold plasma is (inversely) proportional to the $\bar{q}$ on the particle holding the $\bar{q}/\bar{m}$ ratio fixed.
To derive an equation for PIC simulations we use the fact that $\bar{q}/e = n_0 \Delta^2/N$ where $\Delta$ is the cell size (we assume equal cell sizes along the different dimensions) and $N$ is the number of simulation macro-particles per cell, and obtain
\begin{equation} 
\frac{d\gamma_{2D}}{d\omega_p t}=-\frac{1}{4}  \frac{\omega_p^2}{c^2} \frac{\Delta^2}{N}C.
\label{eq:elosspic}
\end{equation}
Equation \eqref{eq:elosspic} demonstrates that for large cell sizes $N$ must be increased to reduce the macro-particle stopping to a desired level.
The factor $C$ is a term that accounts for finite size particle effects which will depend on the details of the particle shape and smoothing used in the simulation. 

Next we consider finite size particle effects and the factor $C$.
These effects will reduce the stopping power of the test particle.
The use of finite size particles reduces both the electric field made by the test charge and the force from this electric field on the particle.
We first discuss the reduction of the electric field. 
The effect of the finite size of the particle on the electric field is found from a convolution of the point particle electric field with the shape factor as a function of grid size

\begin{equation} 
\bar{E}_z(r_{\perp},\xi)= \int_{-\infty}^{\infty}dr_{\perp}'  \int_{-\infty}^{\infty}d\xi' E_z(r_{\perp}-r_{\perp}',\xi-\xi') S(r_{\perp}',\xi'),
\label{eq:convo}
\end{equation}
with $\xi = (z-ct)$.
For linear splines (area weighting) the shape factor is given by
\begin{equation} 
S_1(x)= \frac{1}{\Delta} \left( (1-\frac{x}{\Delta})\eta(x)\eta(\Delta-x)+(1+\frac{x}{\Delta})\eta(-x)\eta(\Delta + x) \right),
 \label{eq:shape}
\end{equation}
where $x$ is the particle position and $\Delta$ is the cell size in a specified direction.
As noted above we consider square cells in 2D but the arguments presented can be easily extended to rectangular cells.
The Fourier transform of this shape function, which is an order 1 spline, is a sinc function squared.
Higher order spline shapes are defined as the convolution of the previous order with the zeroth order shape (nearest grid interpolation),
\begin{equation} 
S_n (x) = \int_{-\infty}^{\infty}dx'  S_0(x-x') S_{n-1} (x').
 \label{eq:splines}
\end{equation}
Therefore, for splines of order n the Fourier transform of the shape function is given by
\begin{equation} 
S(k)_{n}=  \left(\frac{\mathrm{sin}(\frac{k\Delta}{2})}{\frac{k\Delta}{2}}\right)^{n+1}.
 \label{eq:kshape}
\end{equation}
The shape factor is dependent on the cell size and on the interpolation order.
In addition, digital filters are also commonly applied.
A filter of particular interest used for current smoothing is a triangular (1-2-1) filter applied multiple times in each direction.
A compensator \cite{birdsall} is often used to flatten the profile in $k$-space for low $k$.
These filters take the $k$-space form
\begin{equation} 
F(k)_{n}=  \mathrm{cos}^{2n}\left(\frac{k\Delta}{2}\right)
 \label{eq:kfilter}
\end{equation}
where here n is the number passes of the triangular filter.
When the compensator is included the effective filter takes on a more complex expression,
but there is no $k$ dependence in the denominator so smoothing (including compensation) is never as effective as using higher-order splines at reducing aliased noise (the source of particle heating in PIC codes). 
 
\begin{figure}
\centering
\includegraphics[width=3.3in]{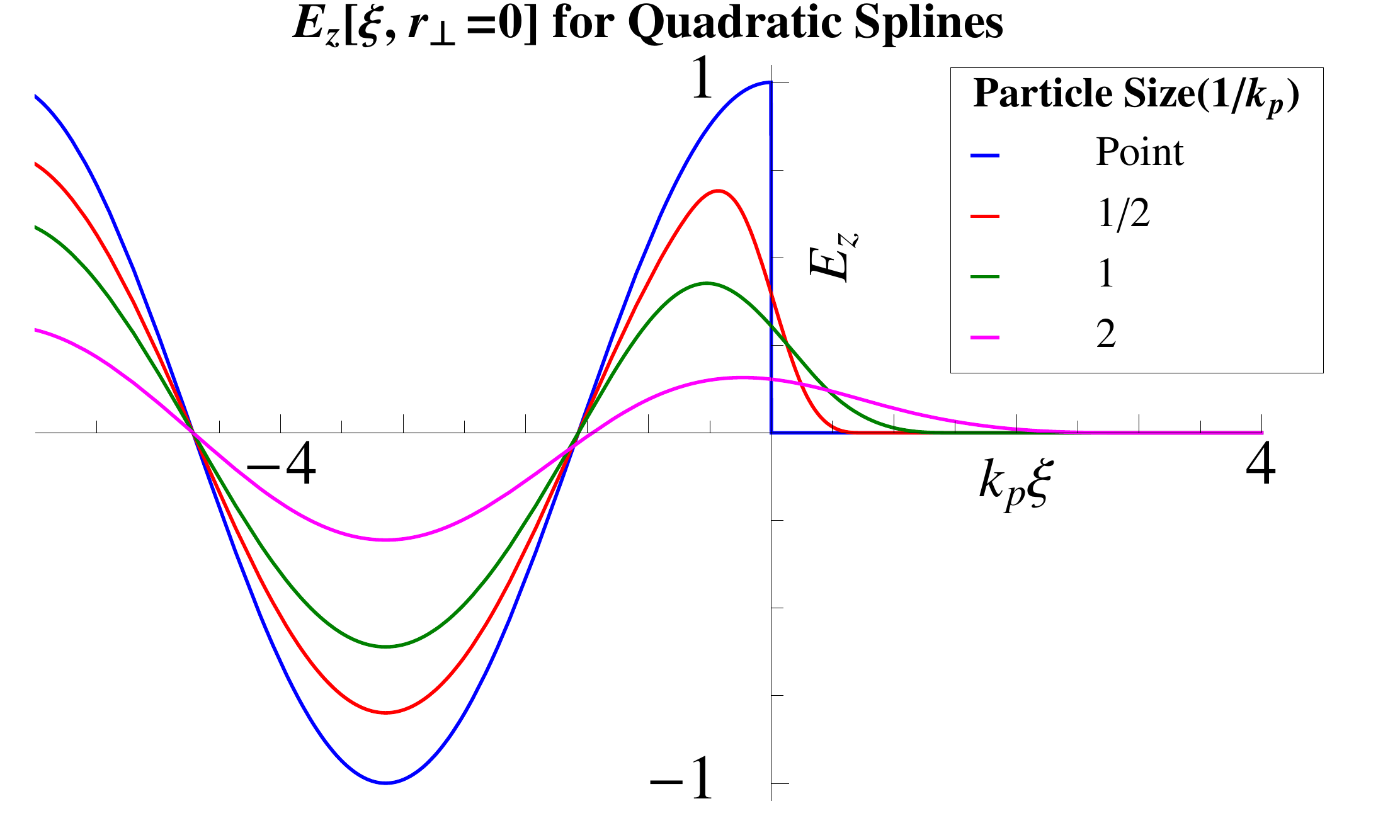}
\includegraphics[width=3.3in]{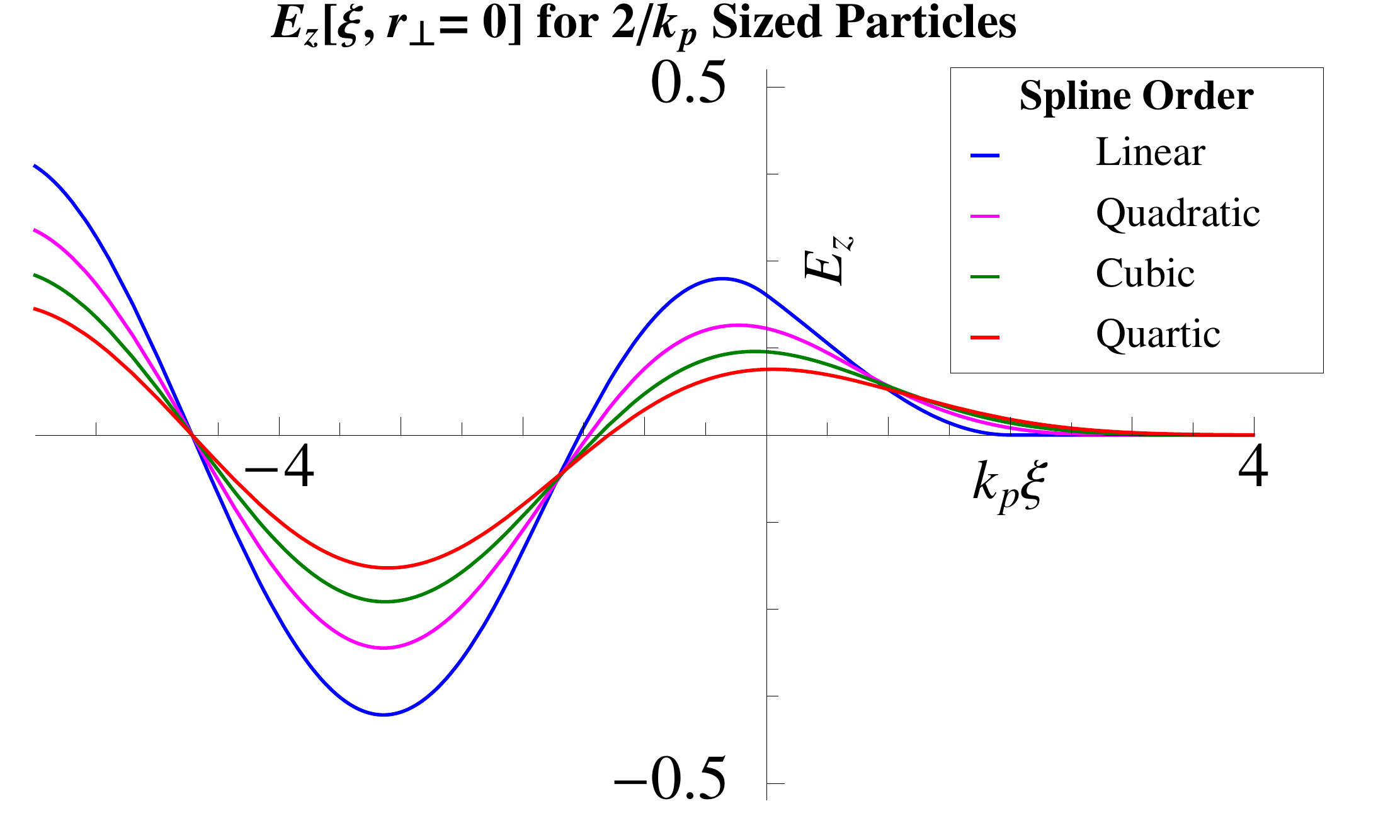}
\caption{\label{fig:fig4}  (Color online) Theoretical curves for the magnitude of the $z$ component of the on axis electric field of a relativistic particle moving in the $z$ direction. a) The strength of the electric field is dependent on cell size with larger cell sizes producing weaker fields. b) The strength of the electric field is also dependent on the interpolation order with higher order interpolation schemes producing weaker fields. 
}
\end{figure}

For illustrative purposes, we examine the electric field generated by different particle shapes where the charge on the particle is kept fixed.
Figure \ref{fig:fig4} shows the magnitude of the axial ($z$) component of the electric field produced on axis ($r_{\perp}=0$) as a function of $z$ by a 2D relativistic particle of finite size  (square) in a plasma according to equation \eqref{eq:convo}.
In figure \ref{fig:fig4}a the particle shape is a second order spline, i.e., $n = 2$ in Eq. \eqref{eq:kshape}, and the cell size is varied from a point particle, to $0.5 c/\omega_p$, to $1.0 c/\omega_p$, and to $2.0 c/\omega_p$ (blue, red, green and magenta curves respectively).
These curves confirm that increasing the cell size for a fixed charge reduces the electric field strength and smooths it out over the location of the particle.
In Figure \ref{fig:fig4}b, we show that increasing the interpolation order while keeping the cell size and charge on the particle fixed also decreases the amplitude of the electric field.
We plot the electric field for  linear (blue), quadratic (magenta), cubic(green), and quartic (red) interpolation while holding cell size fixed at $2.0 c/\omega_p$.
This shows that each higher order spline corresponds to a `wider' particle.
Varying the cell size or varying the particle order both effectively change the size of the particle, however, the use of larger cells also changes the accuracy of the field solver for how light waves propagating in vacuum. 
We also note that although the macro-particle stopping effects are reduced as the cell size is increased to values larger than a skin depth (and Debye length), other issues such as significant modifications of the dispersion properties of both plasma and electromagnetic waves will become important.

Next, we turn our attention from the electric field of the wake to the stopping power of the relativistic particles.
The stopping power is obtained from an integral of the electric field in Eq. \eqref{eq:convo} with the shape function
\begin{equation} 
F_{stopping}= q \int_{-\infty}^{\infty}d\vec {r_{\perp}'}  \int_{-\infty}^{\infty}d\xi' \bar{E_z}(\vec{r_{\perp}'},\xi') S(\vec{r_{\perp}'},\xi').
\label{eq:Force}
\end{equation}
It is worth noting that total force on the particle can also be obtained by integrating the shape function and electric field in k space, 
\begin{equation} 
F_{stopping}= q \int_{-\infty}^{\infty}d\vec{k} S^2(\vec{k}) {E_z}(\vec{k}),
\label{eq:Forceink}
\end{equation}
where through Eq. \eqref{eq:shape} it can be seen how the use of progressively higher order splines (and smoothing) can reduce the stopping force.
The shape functions can be different in each direction, but it is generally a separable function in the each coordinate in which only the cell size varies.
For separable shape functions, it is easy to extend eqs. \ref {eq:Force} and \ref {eq:Forceink} to multi-dimensions.
For simplicity, in what follows we assume the cell sizes are squares in 2D.

\begin{figure}
\centering
\includegraphics[width=3.3in]{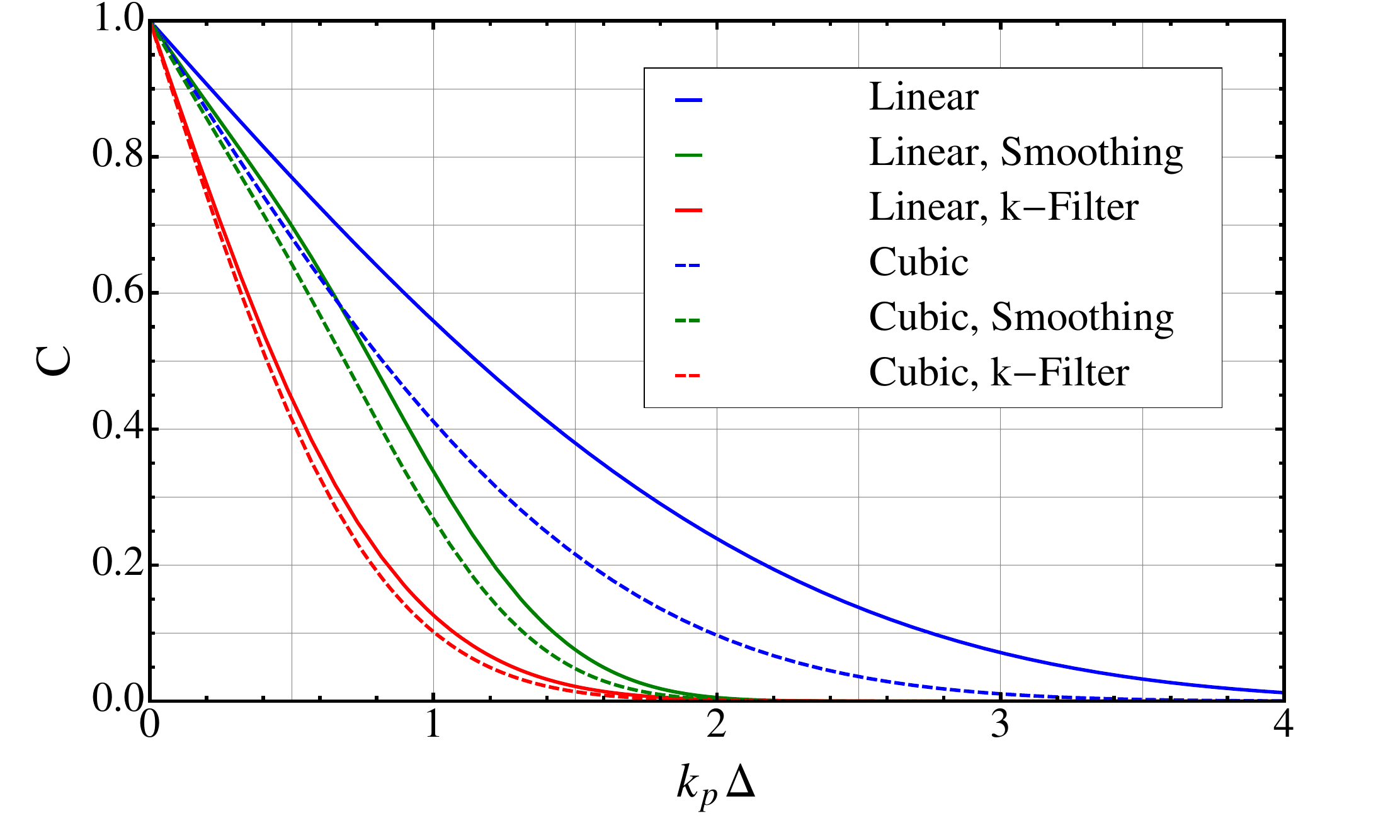}
\caption{\label{fig:fig1}
(Color online)
Reduction in stopping force due to finite-sized particles
(``C" in eq. \eqref{eq:elosspic}.)
`Smoothing' is used in finite-difference codes like Osiris, and here refers to a 4-pass (1,2,1) filter followed by a (-5,14,-5) compensator \cite{birdsall}.
`Filtering' is done directly in k-space and therefore is used in spectral codes like Parsec;
here the filter is $e^{\frac{-k^2 \Delta^2}{2}}$, where $\Delta$ is the cell size (applied to both the particles and the fields).
From the figure it can be seen that smoothing is more effective at reducing stopping than higher-order particle shapes,
and filtering is more effective than smoothing.
}
\end{figure}

It is difficult to obtain analytical expressions for $F_{stopping}$ for the particle shapes used in PIC codes, therefore we simply plot numerical solutions to eq. \eqref{eq:Force}. 
Figure \ref{fig:fig1} shows the effect of varying the spline order and current smoothing on the stopping force.
We plot the force on the particle for linear and cubic splines,
and the same using 4-pass smoothing (which we often use in Osiris) and k-filtering (used in Parsec);
it is normalized to the stopping for a point particle, and therefore this curve is equal to the factor C for two dimensions, $C_2$, in Eq. (\ref{eq:elosspic}).
As stated earlier, both smoothing and higher-order interpolation schemes reduce the macro-particle stopping.
However, for fixed charge per particle, the reduction is only significant for large cell sizes ($\Delta> 1/k_p$ ),
where important skin-depth physics can be missed and where significant numerical dispersion effects can be present \cite{birdsall}.
In essence by increasing the cell size, one is smoothing out the wake field, whose scale length is on order of the skin depth.
For example, for a cell size of $ \nicefrac{1}{2} \, k_p^{-1}$, without current smoothing,
the stopping power is 77\% and 68\% of the point-particle limit for linear and cubic interpolation respectively.
If the cell size is increased to $ \nicefrac{3}{2}  \, k_p^{-1}$, then these values become 37\% and 21\% respectively.
Including current smoothing further reduces the value of $C$ to 0.09 in the linear case and 0.05 in the cubic one.
We note that there are numerical limits to both increasing cell size and increasing interpolation order for the purpose of reducing macro-particle stopping.
For example, the use of large cells can lead to plasma waves and light waves having negative group velocity.
In finite-difference PIC codes where numerical errors cause the phase velocity of electromagnetic waves on the grid to be less then the speed of light, relativistic particles can excite numerical Cherenkov noise which is increased as the time step is reduced below the electromagnetic Courant condition for stability \cite{birdsall}.
In addition, since the hard stability limit for the time step in a plasma is $2/\omega_p$ \cite{birdsall} this introduces a practical limit on the cell size of $2\sqrt{2}  c/\omega_p$ when dealing with relativistic particles.
As seen in equation \eqref{eq:elosspic} the stopping power can be reduced by also increasing the number of particles per cell (or effectively decreasing the particle charge).
This is the best method with respect to not modifying the physics, but it is obviously the most computationally costly method.

We close this section with a brief discussion on macro-particle stopping in 1D and 3D.
The electric field for a infinitesimally thin sheet of charge was give in Eq. \eqref{eq:Ez1D}.
We note that it could also be obtained from integrating the 3D result, Eq. \eqref{eq:Ez}, over a beam with a charge density that is uniform across the beam in the transverse direction.
Interestingly the 1D result is independent of the plasma density, however, the result is only valid if the electric field is still sufficiently small for linear theory to be valid.
This is true if the normalized field value is small,  $\frac{e E_z}{m c \omega_p} \ll 1$ \cite{mori:1988,katsouleas:1987}.
For a macro-particle one needs to integrate Eq. \eqref{eq:Ez1D} over the shape of the particle.
The energy loss for a point (sheet of charge in 1D) particle is given by
\begin{equation} 
\frac{d\epsilon_{1D}}{dt}=- 2 \pi  \bar{\bar{q}}^2 c,
\label{eq:eloss1d}
\end{equation}
which can be written for the Lorentz factor of the test charge as
\begin{equation} 
\frac{d\gamma_{1D}}{d\omega_pt}=- 2 \pi\frac{\bar{\bar{q}}^2}{\bar{\bar{m}} \omega_p c}.
 \label{eq:gloss1d}
\end{equation}
Taking into account the fact that $\bar{\bar{q}}/e = n_0 \Delta/N$ in 1D, we obtain the energy loss in 1D PIC simulations
\begin{equation} 
\frac{d\gamma_{1D}}{d\omega_p t}=-\frac{1}{2}  \frac{\omega_p}{c} \frac{\Delta}{N}C_1
\label{eq:elosspic1d}
\end{equation}
where $C_1$ is the effect of the particle shape in 1D.

In 3D the Green's function logarithmically diverges as $|\vec r_{\perp}|\equiv r$ approaches zero.
However, for a finite size particle the divergence disappears.
A detailed analysis is beyond the scope of this work.
However, in \cite{lu:2005} the result for flat top and Gaussian particle (beam) shapes in the transverse coordinate and Gaussian shapes in the longitudinal directions were given.
Here we summarize the results for Gaussian shaped macro-particles, for which the charge is given by $\frac{q}{(2\pi)^{3/2}}e^{-(z-ct)^2/2\sigma_z^2}e^{-r^2/2\sigma_r^2}$, where $\sigma_z$ and $\sigma_r$ are the particle size in the longitudinal and transverse directions respectively.
For such a charge the amplitude of the wakefield is given in \cite{lu:2005} as
\begin{equation} 
E_{z\,3D} = -q k_p^2 e^{-k_p^2 \sigma_z^2/2} e^{k_p^2 \sigma_r^2/2} \Gamma \left(0,\frac{k_p^2 \sigma_r^2}{2} \right),
 \label{eq:ez3d}
\end{equation}
where $\Gamma (\alpha,\beta) = \int_\beta^\infty{s^{\alpha-1}e^{-s} ds}$ is the incomplete gamma function.
For symmetric macro-particles, $\sigma_r = \sigma_z = \sigma$, we have simply
\begin{equation} 
E_{z\,3D} = -q k_p^2  \Gamma \left(0,\frac{k_p^2 \sigma^2}{2} \right).
 \label{eq:ez3ds}
\end{equation}
We note that in the limit that $k_p \sigma$ approaches zero, $\Gamma(0,\frac{k_p^2 \sigma^2}{2}) \approx ln({1.12}/{k_p\sigma})$, and the electric field on axis reduces to 
\begin{equation} 
E_{z\,3D} = -q k_p^2 ln\left(\frac{1.12}{k_p \sigma}\right).
\label{eq:ez3dlimit}
\end{equation}
This is the peak amplitude of the wake.
To derive how the electric field is distributed over the particle requires carrying out the full integrals over the particle which is beyond the scope of this work.
However, this expression is useful for estimating the macro-particle stopping and we will use it in what follows. 

Following the same procedure as in the 1D and 2D cases, the energy loss in 3D is then described by
\begin{equation} 
\frac{d\epsilon_{3D}}{dt}=- q^2 \frac{\omega_p^2}{c} ln\left(\frac{1.12}{k_p \sigma}\right),
\label{eq:eloss3d}
\end{equation}
or, for the Lorentz factor of the test charge, as
\begin{equation} 
\frac{d\gamma_{3D}}{d\omega_pt}=- \frac{q^2}{m} \frac{\omega_p}{c^3} ln\left(\frac{1.12}{k_p \sigma}\right).
 \label{eq:gloss3d}
\end{equation}
If we assume cubic cell shapes with the same cell size, $\Delta$ in each direction then $q/e = n_0 \Delta^3/N$ and we the cell size for the particle size, $\sigma$,   then we can estimate the macro-particle stopping in 3D, which is given by
\begin{equation} 
\frac{d\gamma_{3D}}{d\omega_p t}=-\frac{1}{4 \pi}  \frac{\omega_p^3}{c^3} \frac{\Delta^3}{N} ln\left(\frac{1.12}{k_p \Delta}\right) C_3.
\label{eq:elosspic3d}
\end{equation}
where $C_3$ is a factor that accounts for additional macro-particle effects. 
We note that for a fixed number of particles per cell, $N$, and a cell size $\Delta \leq c/\omega_p$ (neglecting the exact details of the particle shape), the macro-particle stopping decreases with increasing dimensionality.
Therefore, it requires a larger number of particles per cell in 1D and 2D to control this effect when compared with 3D.
We note that the formalism that has been presented can be used to carry out an exact and detailed analysis of the 3D case as we did for the 2D case.

\section{\label{sec3}  Single Particle Stopping in Fast Ignition Simulations}

We can now make predictions for the stopping of relativistic macro-particles in background plasmas.
Here we focus on the stopping distance in two dimensional fast ignition relevant simulations.
To motivate the importance of macro-particle stopping for fast ignition studies, let us take as example the parameters used in integrated PIC simulations of fast ignition \cite{tonge:2009,chrisman:2008,kemp:2010} aimed at modeling both laser absorption and transport of fast electrons in a background plasma.
We compute the stopping power for the simulations with different resolutions and numbers of particles per cell.
In these simulations an intense laser with a 1 $\mu$m wavelength interacts with a plasma with a nominal density around $100 n_c$ (where $n_c$ is the critical plasma density for the laser).
The number of macro-particles per skin depth squared (PPSD$^2$) in these simulations varied between 6 and 24, the cell sizes varied from 0.5 to 2 $k_p^{-1}$, and the particle order varied from linear to cubic.
Despite these differences, the predicted energy loss rate was similar and it only varied in the range $0.12 - 0.34$ MeV$/\mu$m.
The simulations used different plasma lengths so the single particle energy loss across the simulation box varied between 4 and 18 MeV.
For example, in the paper by Tonge et al.\cite{tonge:2009} the cell size was 0.5 $c/\omega_p$, with 4 particles per cell (N = 4) giving 16 PPSD$^2$, and quadratic splines and 5-pass compensated smoothing were used, leading to a stopping force of 0.011 $m_e c \, \omega_p$, or an energy loss rate of 0.337 MeV$/\mu$m at a density of $100 n_c$.
In each of these cases the energy loss of a macro-particle due to enhanced stopping is comparable or larger than the typical fast-electron energies relevant for fast ignition.
Therefore the enhanced stopping of macro-particles could greatly affect the results in each case.

We verify these predictions by running simulations of single particle stopping with the PIC codes OSIRIS \cite{fonseca:2002,fonseca:2008,fonseca:2013} and PARSEC \cite{parsec}.
The first is a finite difference code, whereas the second is a spectral code which will not produce numerical Cherenkov radiation of a single particle without the effect of aliasing (note that there is radiation from aliasing effects).
In OSIRIS we also use current smoothing and compensation, while in PARSEC the particles also have a Gaussian shape with a size equal to the grid size.
For the single particle stopping tests, we run 2D simulations with periodic boundary conditions.
In these simulations the cell size is 0.5 $c/\omega_p$, the box size is 128 $c/\omega_p$ $\times$ 128 $c/\omega_p$ and quadratic splines are used for interpolation.
Both ions and electrons are included, the mass ratio being 3672 (Deuterium plasma).
Simulations where performed with N = 64 in the background plasma,
the background temperature is 100eV,
and the duration of the simulations is 44.8 $\omega_p^{-1}$.
In order to study the wake driven by a single relativistic electron in the background plasma we use a subtraction technique \cite{decyk:1987,grismayer:2011}, where two identical simulations are run, but in one of them a single 50 MeV electron moving in the $z$ direction (hereafter, $x_1$ direction) is added.
The wake electric field is obtained by subtracting the electric fields of both simulations.
This tracer particle has the same $\bar{q}$ as particle with 16 particles per $1/k_p^2$.

\begin{figure}
\centering
\includegraphics[width=3.3in]{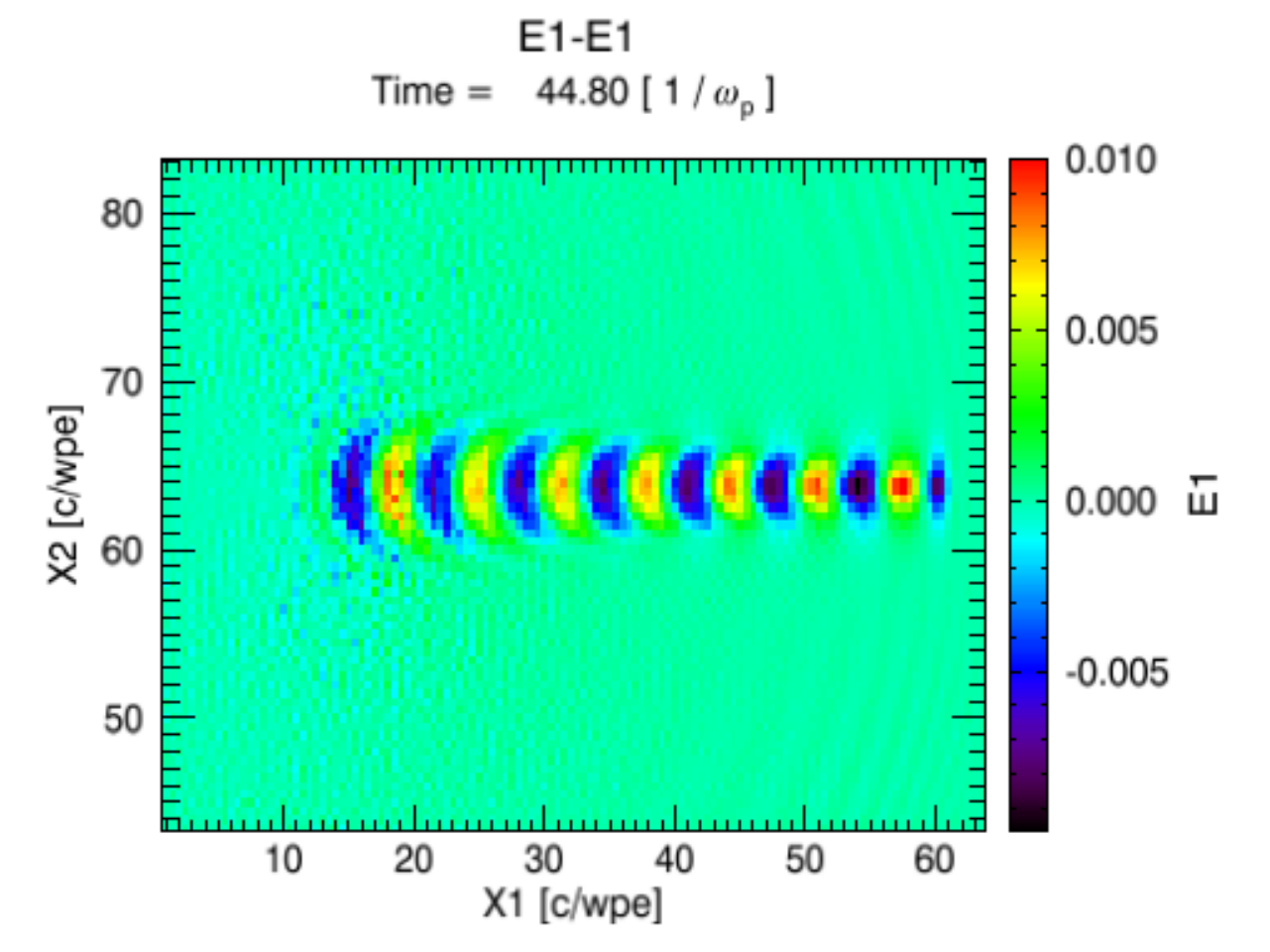}
\includegraphics[width=3.3in]{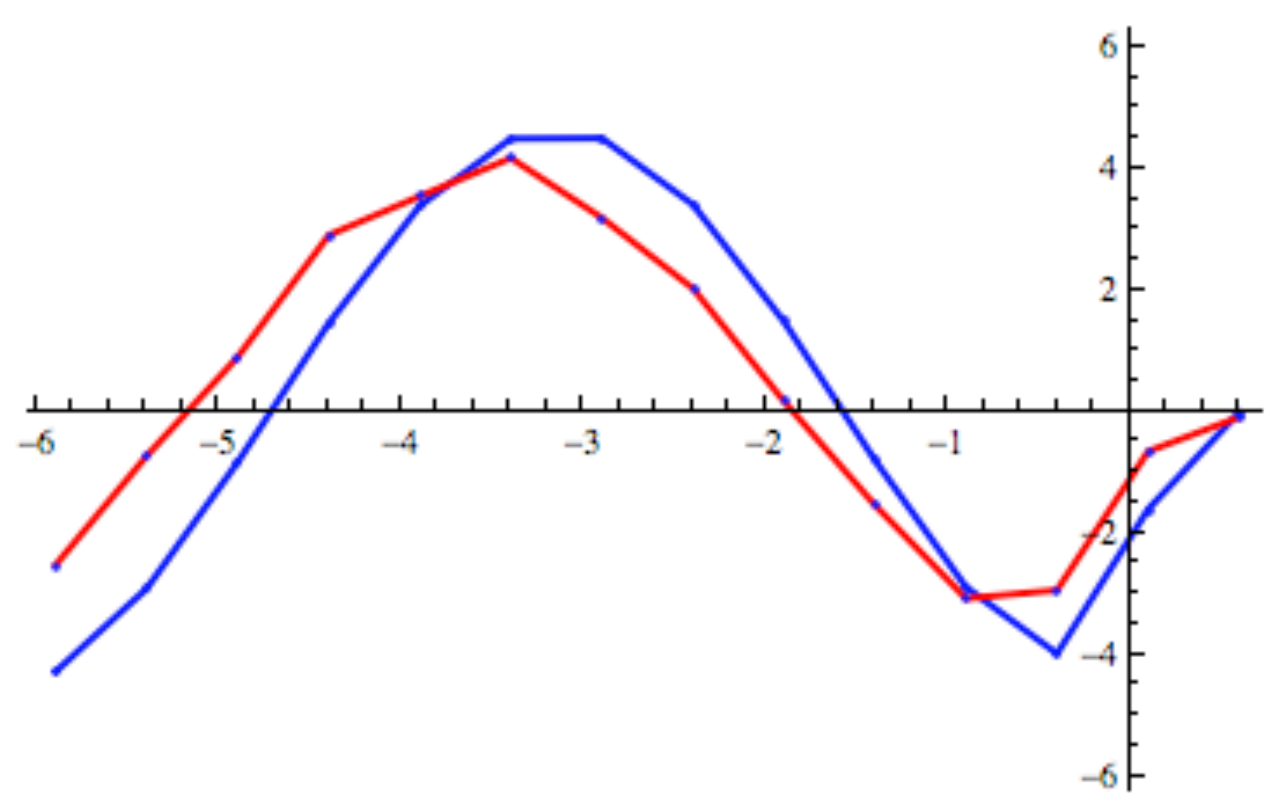}
\caption{\label{fig:fig2}  (Color online) a) Wake formed in the $E_1$ field by a relativistic particle traveling along the $x_1$ direction in a 2D PIC simulation. This wake is predicted  by wakefield theory. b) A comparison of the lineout along $x_1$ of the $E_1$ field near the relativistic particle predicted by theory (blue) and from the PIC simulation (red). The curves are in close agreement with the simulation curve (red) having a smaller amplitude and being slightly offset.
}
\end{figure}

Figure \ref{fig:fig2}a shows the $E_1$ obtained using the subtraction technique from PARSEC simulations, thereby eliminating the noise and clearly showing the wake of the fast particle.
In figure \ref{fig:fig2}b we show a zoomed-in line-out on axis of the wake field from \ref{fig:fig2}a (red),
along with the theoretical prediction from equation \eqref{eq:Ez} (blue).
As can be seen the results are in good agreement,
although the simulation does produce a more ragged and slightly weaker field than predicted by theory, even with the finite sampling taken into account.

We also compared the energy loss observed in the simulations to the theory presented in Section \ref{sec2}.
According to equation Eq.$\,$(\ref{eq:elosspic}) a point particle will lose an energy of 0.7 $m_e c^2$ in 44.8 $\omega_p^{-1}$;
including finite-size effects this is decreased to 0.30 $m_e c^2$ in PARSEC and 0.47 $m_e c^2$ in OSIRIS.
In our simulations the tracer particle lost 0.44 $m_e c^2$ in PARSEC and 0.64 $m_e c^2$ in OSIRIS.
Both results are on the order of and less than the point particle result, and the ratio of the two values is correct to theory.
That both results show weaker finite-particle effects than predicted by macro-particle theory may be due to the background plasma being discrete and warm.
Understanding these differences is an area for future work.
 
We have also performed a series of test particle simulations with PARSEC to examine the effect of background temperature and particle energy.
Recall that the theory assumes a cold fluid plasma and that the particle is highly relativistic.
These 2D PIC simulations use the same simulation box and numerical parameters of the previous simulations, but we varied the energy of the relativistic electron ($> 1$  MeV) and of the background plasma (100 eV - 1MeV).
For background temperatures up to 10keV and all fast electron energies examined, the energy loss is relatively insensitive to either parameter, and is $\sim 14\%$ less than the theoretical value.
For background temperatures $> 100$ keV, the velocity of background particles becomes relativistic, the assumption of a cold background plasma does not hold and the simulation results deviate more strongly.

We have tested $N =$ 4, 16, 64, and 256 for the background plasma and the differences to the stopping of a single particle were negligible for background temperatures below 10keV.
We therefore conclude that equation \eqref{eq:elosspic} is useful for all initial energies above 1MeV and for background temperatures up to $\sim$ 10keV.

\section{\label{sec4}  Relaxation of high-energy tails}
\begin{figure}
\centering
\includegraphics[width=3.3in]{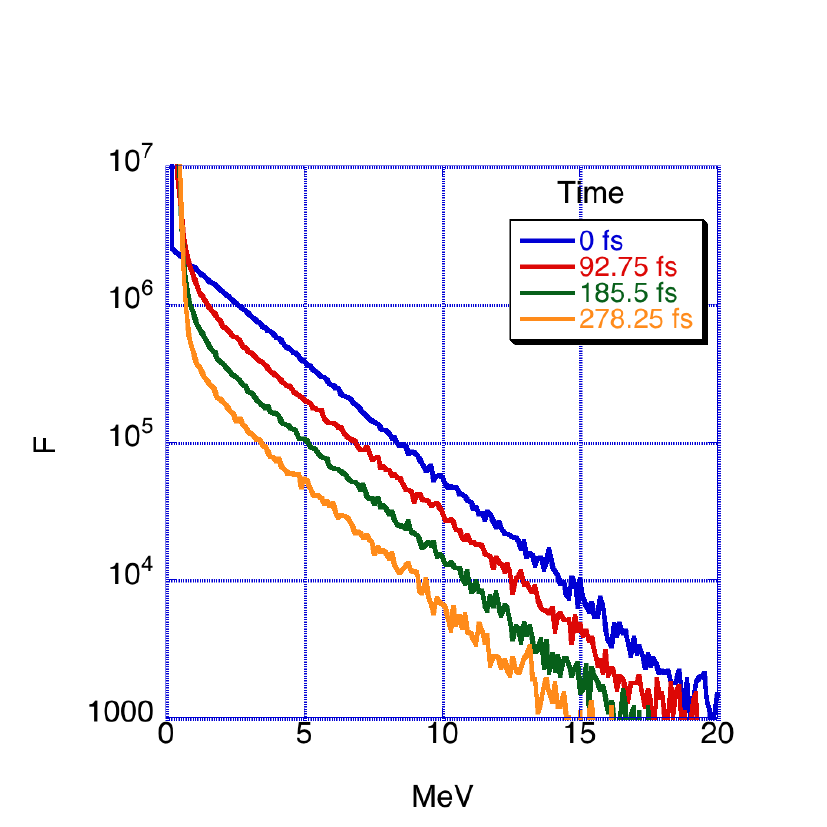}
\caption{\label{fig:fig6}  (Color online) Relaxation of high-energy tails in a periodic simulation of a cold background and a hot electron tail. The blue curve shows the initial energy distribution, with the red, green and yellow curves showing the relaxation of the distribution with time, 92.75 fs apart when scaled to a density of 100$n_c$.  The curves show that the high-energy particles uniformly loose energy as predicted by theory.}
\end{figure}

We now investigate the macro-particle stopping for a distribution of fast electrons with a high-energy tail.
These simulations are run with PARSEC and model a spatially uniform cold (1 keV) background plasma with a relativistic tail distribution with a slope temperature of 2.5 MeV and with 4$\%$ of the particles in the tail.
The simulations have periodic boundaries, with a box size of 256 cells by 256 cells, a cell size of 0.5$c/\omega_p$, and $N = 16$ or 64 PPSD$^2$.
Second order particle interpolation is used with Gaussian shaped particles with particle sizes equal to the grid size.
In Figure \ref{fig:fig6} the blue curve shows the initial energy distribution, with the red, green and yellow curves showing the relaxation of the distribution with time, 92.75 fs apart when scaled to a background density of $10^{23}$ cm$^{-3}$.
The curves show that the high-energy particles uniformly loose energy as predicted by theory.
The predicted energy loss is 0.09 MeV$/\mu$m (0.027 MeV/fs) and the simulations show an energy loss of 0.06 MeV$/\mu$m, which is less than both theory and single particle simulations.

In single particle simulations the energy loss is 86$\%$ of theory and insensitive to background particle count;
in simulations with a high-energy tail the discrepancy with theory is dependent on the PPSD$^2$ because this effects the distribution of particles in the tail.
For simulations where we vary the PPSD$^2$ but keep other parameters constant,
we found energy loss of 71$\%$, 66$\%$, and 53$\%$ of the predicted value for 16, 64, and 256 PPSD$^2$ respectively.
This effect is likely due to the more even distribution of particles in the tail across cell boundaries as $N$ (not necessarily PPSD$^2$) increases.
For the particles in the tail the number of particles per cell varied from $N = 0.16$, 0.64, and 2.56 respectively (recall that $4\%$ of the particles are in the tail).
As the tail becomes more uniformly distributed in space the discrete effect of macro-particles is decreased.
In simulations relevant to fast ignition the electrons are accelerated by the laser in bunches \cite{may:2011} separated by half a laser wavelength which is comparable to a plasma wake wavelength.
Therefore, the particles in these bunches will not interact through individual wakes but rather through collective wakes. 
If the bunch duration is short compared to a skin depth the wakes will collectively add leading to enhanced stopping while if the particles are bunched on scales much longer than the skin depth then the stopping power will be comparable to the single particle stopping power. 

\section{\label{sec5}  Integrated Fast Ignition Simulations}
\begin{figure}
\centering
\includegraphics[width=3.3in]{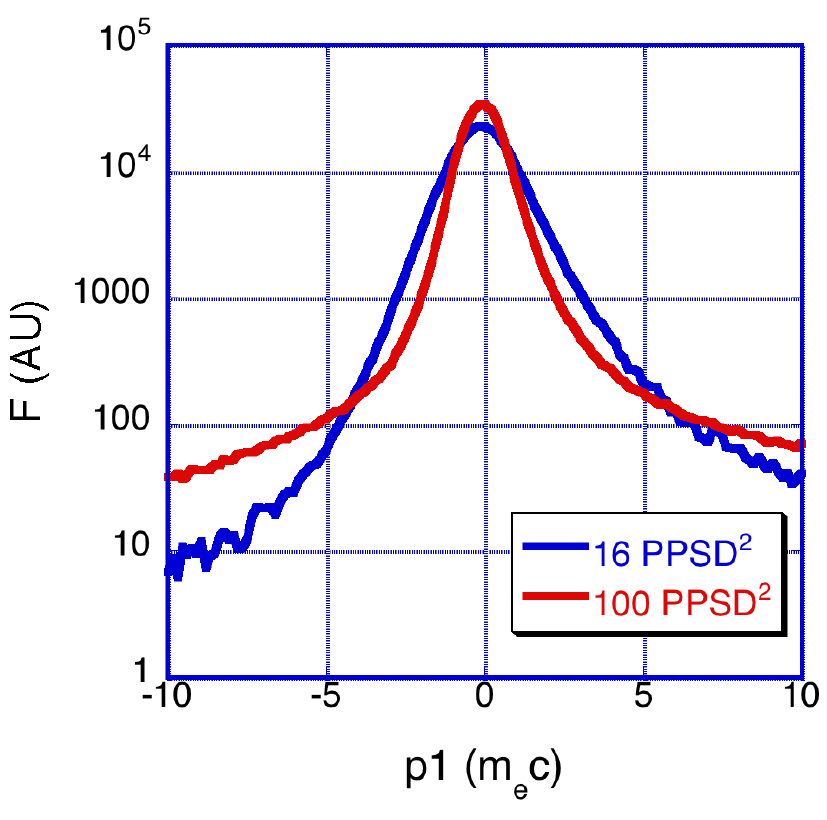}
\includegraphics[width=3.3in]{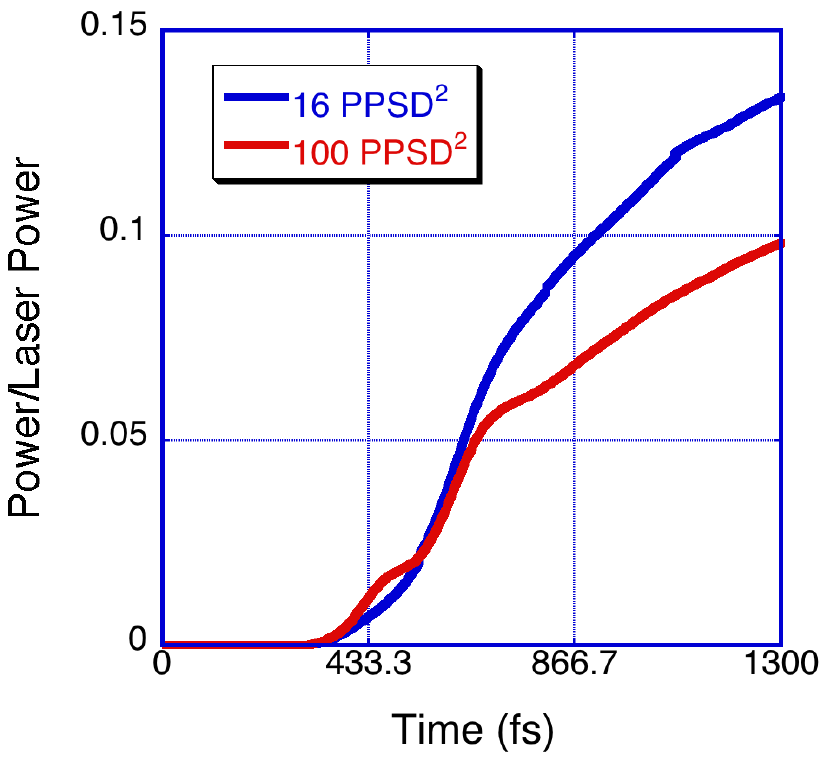}
\caption{\label{fig:fig8}  (Color online) Panel a) shows the distribution function for forward momentum ($p_1$)  3 $\mu$m in front of the target core at 835 fs for 16PPSD$^2$ and 100 PPSD$^2$ simulations respectively. These show that macro-particle stopping results in less heating of the plasma. Panel b) compares the power delivered to the core for the 16PPSD$^2$ and 100 PPSD$^2$ simulations. These show that greater macro-particle stopping also results in lower energy electrons carrying energy to the target core.
}
\end{figure}

We will now revisit the integrated simulations using isolated fast ignition targets from Tonge et al. \cite{tonge:2009}.
In these simulations a 50$\mu$m radius 100$n_c$ target with a 20$\mu$m core region is illuminated with an $I=8\times10^{20}$W/cm$^2$, $\lambda = 1 \mu$m laser with a 20$\mu$m spot size;
detailed parameters are given in Tonge et al.\cite{tonge:2009}.
Here we reproduce these simulations using cubic splines (quadratic splines were used in the original simulations) and
varying the PPSD$^2$ while keeping all other parameters fixed.
By increasing the PPSD$^2$ from 16 (original value) to 100 we decrease the stopping by 6.25 times and we observe a significant change in the dynamics of the simulation due to macro-particle stopping.
The stopping decreases the energy of hot electrons while increasing the heating of the background plasma between the laser-plasma interface and the target core.
Figure \ref{fig:fig8} shows the momentum distribution function in the isolated target,
835 fs after the laser strikes,
at the longitudinal ($x_1$) position 2 $\mu$m behind the laser-plasma interface and integrated along the transverse direction ($x_2$) for the 16 PPSD$^2$ or 100 PPSD$^2$ cases.
The blue curve (16 PPSD$^2$) is clearly wider than the red curve (100 PPSD$^2$) indicating the plasma is hotter in the interaction region.
Panel b) shows the power absorbed in the target core as a function of total laser power.
From the initial heating at 400 fs up to 600 fs the two simulations have similar power delivered to the core,
although the red curve has a bump at 450 fs which is due to higher level of refluxing off the back of the target early in the high particle count simulation.
After 600 fs the curves diverge, with the lower PPSD simulation showing a larger heating of the core;
this is due to both energy of the electrons reaching the core being lower and hence the stopping of fast electrons in the core being enhanced, and to the increased
heating of the background plasma in front of the core that causes heated background electrons to travel into the core and redeposit their absorbed energy.
This overall effect was to increase the power delivered to the core by $25\%$ at 1.3 ps.
In the lower absorption case (higher PPSD case) the laser is still capable of delivering $10\%$ of its power to the core.

Let us now compare the magnitude single particle stopping effect to that seen by a real particle (although these are 2D simulations, the 3D result is equivalent up to a factor of $\mathcal{O} (1)$.)
For a $1 \mu$m laser, $100 n_c = 10^{23}$ cm$^{-3}$, the cell size of $0.05 c/\omega_0$ is $\sim 0.008 \mu$m, giving ~$5\times10^4$ electrons in a cell volume.
For $N = 25$, and taking into consideration finite particle size effects, the stopping is $\sim 1300$ times stronger in the simulation than would be seen for a physical electron.
To make the real and simulated macro-particle stopping powers numerically equivalent at $n = 100 n_c$ and retain the skin depth physics (i.e. cell size $\leq c/\omega_p$) we would have to increase the particle count to $N > 500$ and use current smoothing and cubic splines.
This is impractical for multi-dimensional PIC studies of relevant fast ignition scales.
However, it is not important the make the macro-particle stopping power comparable to the real stopping power, rather it is important to make it small enough over the plasma size so that this enhanced stopping does not change the results.
On the other hand, if one is modeling the heating of the core at very high densities, then since the stopping power of the electrons is the physics process that is most important and it needs to be quantitatively correct.
This is similar to the effect of the collisions between macro particles.
If one is modeling a region of plasma for which collisions are not important it is not important that the collision frequency is correct, rather it must be kept small enough that the growth rates and dispersion properties of collisionless processes are not modified. 

\section{\label{sec6} Particle Splitting}

We now discuss a scheme that can potentially overcome this effect of macro-particle stopping in PIC simulations associated with laser-solid interactions, fast-ignition, and relativistic shocks.
We have shown that reducing the charge on the relativistic, i.e., `beam',  electrons is an effective method to alleviate the enhanced stopping of relativistic particles due to their wakes.
However, in the majority of the simulations of interest we want to study the way particles are accelerated, e.g. by the laser in fast ignition scenarios or by the shock in astrophysical plasmas, and we do not know beforehand which particles will be accelerated (fast) and which will be background plasma particles.
Thus, in a standard PIC simulation reducing the charge of a fast particle requires reducing the initial charge of all simulation particles, leading to a prohibitive particle count.

In many of these scenarios, the super-thermal particles which are stopped by macro-particle effects are only a small percentage of the total particle count - although at the same time these particles can have a profound effect on plasma dynamics.
Under these conditions an algorithm that splits the relativistic electrons into macro particle electrons with less charge might be effective.
We have implemented and experimented with a splitting algorithm that works as follows.
For a given simulation we define the minimum particle energy above which macro-particle stopping will become important and the maximum particle charge that high-energy particles should have for accurate description of their stopping.
In the simplest algorithm these can be fixed values, for instance associated with the initial laser and plasma conditions.  
It would also be possible to make them dynamic values that depend on for instance the local background plasma conditions that vary due to density gradients or plasma steepening.
The simulation is initialized in the standard way, however, after each n time steps (in the future the value of n could change statically or dynamically as the simulation progresses), we calculate the energy and charge of each particle and compare it to the minimum energy and maximum charge defined.
If they are both larger, then we reduce its charge and mass by a factor of two and then duplicate the particle.
This process is repeated until the particle charge is smaller than the defined value.
Thus, we effectively increase the particle count for the high-energy part of the distribution function.
It is important to note that in most cases, because of the wide difference between the background thermal and `fast' particle energies, the efficiency of the splitting is relatively insensitive to the exact energies chosen as the splitting points.
Also, since the number of fast particles is significantly smaller than the number of background particles, the computational overhead of splitting is not dramatic.
We also have found the results are relatively insensitive to the choice of n.

It is important to ensure that the duplicated particles move away from each other.
If the duplicated particles have the exact same position and momentum then they will never move apart and their wakes will add coherent and the stopping power will remain unchanged.
Furthermore, the use of finite size particles means that particles that start off close together within a cell will move apart slowly.
Several schemes can be thought of to separate the duplicated particles, such as randomly shifting the position of the new particle within the same cell, slightly shifting the momentum of the two particles in a way that conserves the total momentum (this can introduce a small divergence), or simply relying on a collisional operator, which is already used in many of the scenarios of interest, to naturally separate the two particles in phase space.
We have experimented both with the momentum shifting scheme and with the use of a Monte-Carlo binary Coulomb collisions operator \cite{takizuka:1977,nambu:1998,sentoku:2008,peano:2009}, with overall satisfactory results. 

We next show results using the OSIRIS Coulomb collisions module for electron-ion collisions.
We initialize a 2D plasma in a box $1 \mu$m wide and periodic in the transverse direction, and $200 \mu$m long in the longitudinal direction.
The box is filled with a $100 n_c$ plasma,
except for a thin ($2 \mu$m) vacuum layer to the left which isolates the plasma from this wall;
a laser with $I = 5\times 10^{19}$ W/cm$^2$ (normalized vector potential $a_0 = 6$) and $\lambda = 1 \mu$m is incident from this direction.
In order to avoid using a larger simulation box to capture the self-consistent plasma expansion due to laser heating of the plasma surface,
we used infinitely heavy ions.
We use a linear density ramp from 0-100 $n_c$ in the first micron of plasma.
We use cells with a size of $0.5 c/\omega_p$,
and either 16 or 512 particles per cell,
corresponding to 64 or 2048 PPSD$^2$, respectively; the time step satisfies the Courant condition almost exactly, to reduce numerical Cherenkov.
We use third order particle shapes, with 4-pass current smoothing with a compensator.

\begin{figure}
\centering
\includegraphics[width=3.3in]{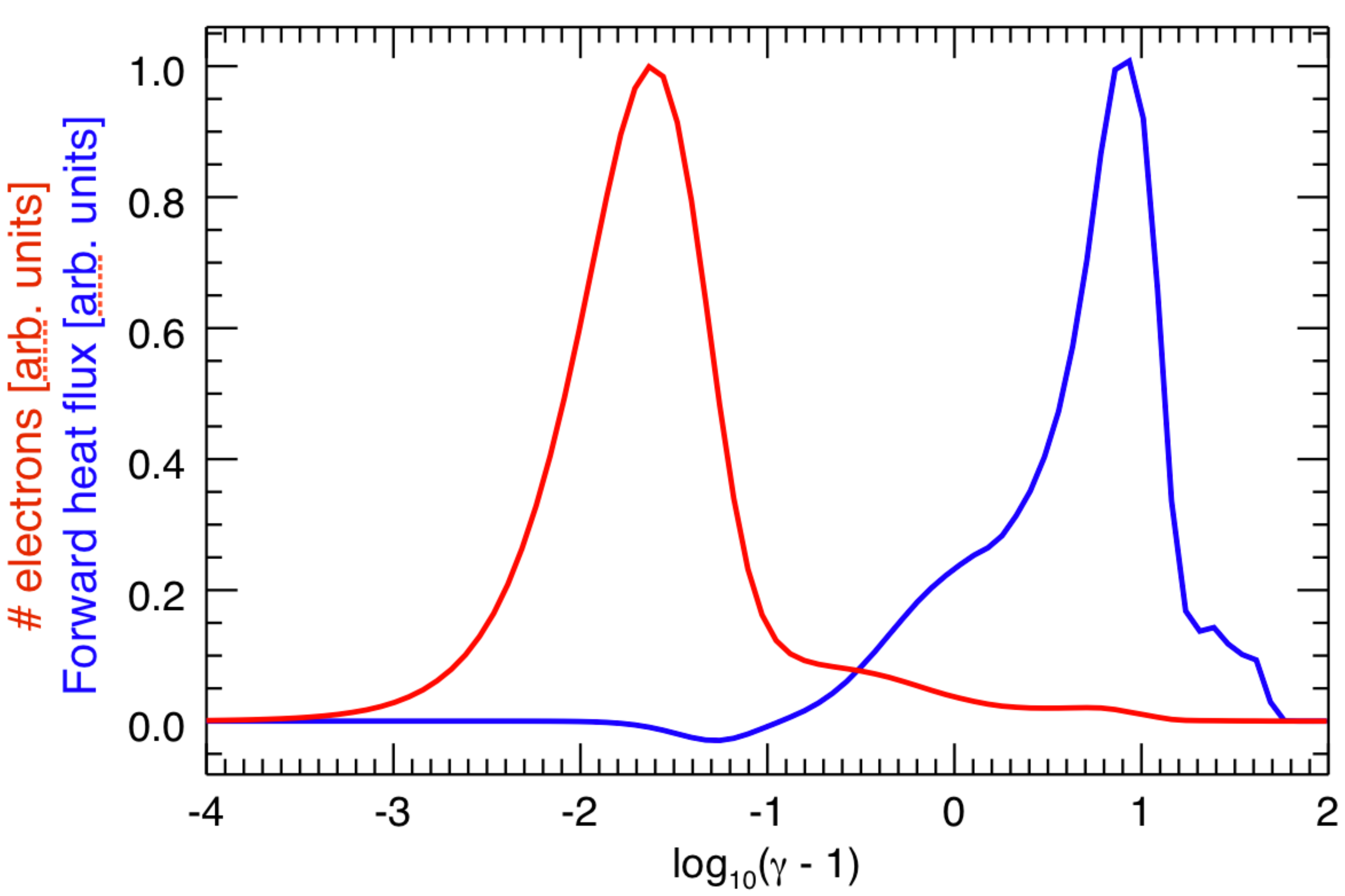}
\caption{\label{fig:count}  (Color online) Particle count vs. the logarithm of the kinetic energy (red), and the same data weighting each particle by the heatflux it carries in the $x_1$ direction (blue.) Arbitrary units, both independently normalized to a peak of 1.}
\end{figure}

\begin{figure}
\centering
\includegraphics[width=6in]{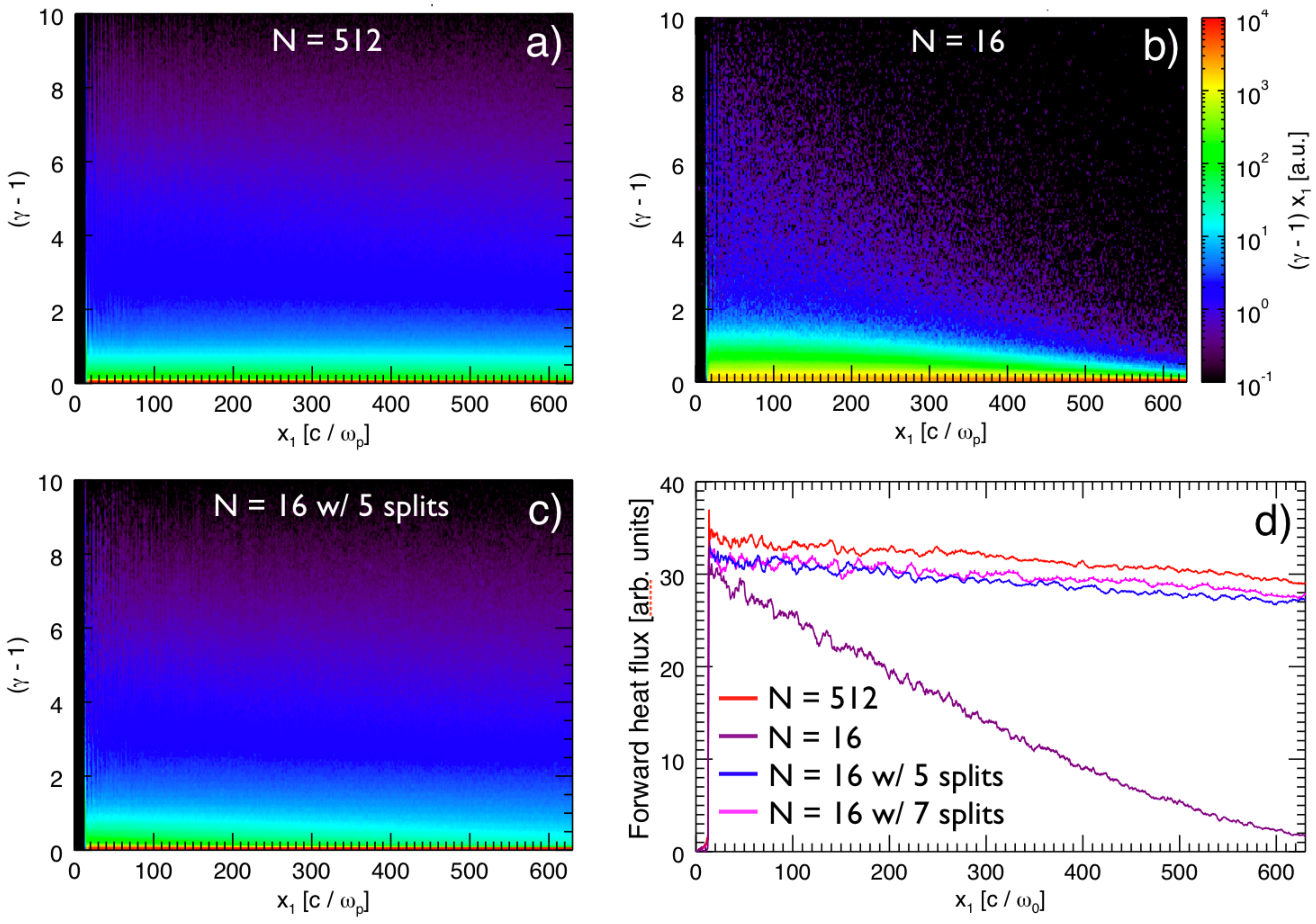}
\caption{\label{fig:splitting}  (Color online) Particle count as a function of $\gamma$ and forward position for a) 2048 PPSD$^2$, b) 64 PPSD$^2$, and c) 64 PPSD$^2$ but with up to 5 particle splits for superthermal particles. d) Forward heatflux as a function of forward position for 2048 PPSD$^2$ (red curve), 64 PPSD$^2$ (magenta), 64 PPSD$^2$ w/5 splits (blue), and 64 PPSD$^2$ w/7 splits (pink). All data in arbitrary units but with equivalent normalization.}
\end{figure}

In figure \ref{fig:count} we show the particle count and forward heat-flux of a typical laser-solid simulation as a function of kinetic energy on a logarithmic scale
for $N=512$ (note that particle count is not corrected for bin size).
The first thing to notice is that the majority of the energy flux is being carried by particles with $\gamma \sim 2-10$,
demonstrating that the energy is being carried by relativistic electrons and that macro-particle stopping needs to be considered. 
It is also clear that the peak of the heat-flux and the peak of the particle count are separated by two orders of magnitude indicating that almost all of the forward heat flux is carried by only a few percent of the particles.
This illustrates that if decreased the particle charge and increased the particle count of only a small region of phase space we could greatly minimize the importance of macro-particle stopping without significant computational expense.
To illustrate the effectiveness of the splitting algorithm, in Figure \ref{fig:splitting} we show the total forward heat flux as a function of longitudinal position
for the  $N = 512$ case as well as a $N = 16$ case with no splitting and a  $N = 16$ case  with five particle splits (for final particle size equivalent to $N = 16 \times 2^5 = 512$).
Figure \ref{fig:splitting}b clearly shows significant loss of beam energy going into the plasma for the $N = 16$ case as compared to the $N=512$ case.
In fig. \ref{fig:splitting}b 
the heatflux essentially vanishing by $x_1 = 100 \mu$m.
Taking the  peak beam particle energy as $2 a_0 m_e c^2$ \cite{may:2011},
we find for these parameters a stopping distance of 97.5 $\mu$m,
consistent with these results.
For the $N = 512$ case (Fig. \ref{fig:splitting}a), we predict a decrease of the particle Lorentz factor, $\gamma$, of 0.375 over the same distance,
which we estimate to be $3-6\%$ of beam particle energy.
The larger apparent loss of heat flux moving forward in the case of $N = 512$ is due to a number of non-macro-particle effects,
including the increase of laser absorption as a function of time combined with the time-of-flight of electrons,
the fact that not all heat flux is being carried by highly-relativistic particles,
and the effects of the wakes of at least partially coherent particle bunches.
We note that collisional stopping, however, is not a significant factor in this regime.
More importantly, the results using $N = 16$ but with 5 binary particle splits (Fig. \ref{fig:splitting}c) are identical to the $N = 512$ simulation.
Figure \ref{fig:splitting}d shows the comparison of the heat flux carried in the forward direction as a function of the longitudinal positions for the  different cases.
Again, it is possible to observe that by splitting the high-energy particles the macro-particle stopping is controlled.
For the example shown, using the splitting algorithm leads to only a  factor of two increase in the computational time, giving a total computational savings factor of 16 to reach the same accuracy in terms of macro-particle stopping.
This illustrates the efficiency of this  algorithm and its usefulness for carrying out multi-dimensional studies of large plasma volumes.
We should also note that using this splitting algorithm can bring important statistical advantages when computing collisions between fast and background particles in many of the scenarios of interest.
We note that that this algorithm can lead to load balancing issues.
OSIRIS does have a dynamic load balancing capability and we will experiment with this as part of future work.

\section{\label{sec7} Summary}
In this paper we have shown that relativistic particles moving in a cold background plasma in PIC simulations are susceptible to enhanced stopping due the use of macro-particles.
The stopping scales as  $q^2/m$ so that particles with large charge but with the correct charge to mass ratio will stop more rapidly.
This stopping is due to the wakefield created by relativistic particles and it can be predicted using wakefield theory developed for studying plasma wakefield acceleration.
We reviewed the derivation of the wakefields created by a point particle moving near the speed of light for one dimension (charge sheet), two dimensions (line of charge), and three dimensions.
We used this wakefield (Green's function) to derive the wakefields created by finite size particles (such as those used in PIC simulations), and also calculated the force on a particle from its own wakefield to get the stopping power.
We also studied how the stopping depends on the cell size, particle shape, and dimensionality.
The enhanced stopping is mitigated through the use of larger cells, higher order particle shapes, and current smoothing as well as with a decrease in the macro-particle charge.
We found good agreement between the theory and results against PIC simulation from a finite difference PIC code (OSIRIS) and a spectral PIC code (PARSEC).
We also studied how a distribution of electrons containing a hot tail relaxes and showed that the macro-particle stopping process dominates how the tail relaxes.
We also reexamined previous results in intense laser-solid interactions, such as the isolated target simulations in \cite{tonge:2009} and found that indeed this effect modified the physics.
When the charge per particle was reduced by a factor of 6.25  the amount of laser energy being deposited in the core decreased by only $25\%$ from the values quoted in \cite{tonge:2009} .
Our expressions also predict that simulations done by others will also be impacted by this effect \cite{chrisman:2008,kemp:2010}.
Lastly, we described a particle splitting algorithm that can reduce this effect by decreasing the charge of high-energy particles as the simulation evolves.
We found that this algorithm can successfully control the macro-particle stopping in PIC simulations of high-energy density scenarios in a computationally efficient way.

\section*{acknowledgments}
We acknowledge a useful discussion with Dr. Max Tabak which was an initial motivation for this work in 2009.
When finalizing this manuscript it came to our attention that independent work on this topic has been recently carried out by
Kato \cite{kato:2013}.
This work was supported by the US Department of Energy under the Fusion Science Center on Extreme States of Matter and Fast Ignition Physics through a subcontract from the University of Rochester, and by contracts  DE-NA0001833,  DE-SC0008316, and DE-SC0008491, and the NSF grant number NSF-ACI-1339893.
This work was also performed under the auspices of the U.S. Department of Energy by Lawrence Livermore National Laboratory under Contract DE-AC52-07NA27344.
FF acknowledges financial support by the LLNL Lawrence Fellowship.
The work of LOS and RAF is supported by the European Research Council (ERC-2010-AdG Grant No. 267841).
The simulations were carried out on the Hoffman and Dawson2 Clusters at UCLA and on Intrepid at Argonne National Laboratory.

\bibliography{stopping}

\begin{thebibliography}{29}%
\makeatletter
\providecommand \@ifxundefined [1]{%
 \@ifx{#1\undefined}
}%
\providecommand \@ifnum [1]{%
 \ifnum #1\expandafter \@firstoftwo
 \else \expandafter \@secondoftwo
 \fi
}%
\providecommand \@ifx [1]{%
 \ifx #1\expandafter \@firstoftwo
 \else \expandafter \@secondoftwo
 \fi
}%
\providecommand \natexlab [1]{#1}%
\providecommand \enquote  [1]{``#1''}%
\providecommand \bibnamefont  [1]{#1}%
\providecommand \bibfnamefont [1]{#1}%
\providecommand \citenamefont [1]{#1}%
\providecommand \href@noop [0]{\@secondoftwo}%
\providecommand \href [0]{\begingroup \@sanitize@url \@href}%
\providecommand \@href[1]{\@@startlink{#1}\@@href}%
\providecommand \@@href[1]{\endgroup#1\@@endlink}%
\providecommand \@sanitize@url [0]{\catcode `\\12\catcode `\$12\catcode
  `\&12\catcode `\#12\catcode `\^12\catcode `\_12\catcode `\%12\relax}%
\providecommand \@@startlink[1]{}%
\providecommand \@@endlink[0]{}%
\providecommand \url  [0]{\begingroup\@sanitize@url \@url }%
\providecommand \@url [1]{\endgroup\@href {#1}{\urlprefix }}%
\providecommand \urlprefix  [0]{URL }%
\providecommand \Eprint [0]{\href }%
\providecommand \doibase [0]{http://dx.doi.org/}%
\providecommand \selectlanguage [0]{\@gobble}%
\providecommand \bibinfo  [0]{\@secondoftwo}%
\providecommand \bibfield  [0]{\@secondoftwo}%
\providecommand \translation [1]{[#1]}%
\providecommand \BibitemOpen [0]{}%
\providecommand \bibitemStop [0]{}%
\providecommand \bibitemNoStop [0]{.\EOS\space}%
\providecommand \EOS [0]{\spacefactor3000\relax}%
\providecommand \BibitemShut  [1]{\csname bibitem#1\endcsname}%
\let\auto@bib@innerbib\@empty
\bibitem [{\citenamefont {Dawson}(1983)}]{dawson:1983}%
  \BibitemOpen
  \bibfield  {author} {\bibinfo {author} {\bibfnamefont {J.~M.}\ \bibnamefont
  {Dawson}},\ }\href {\doibase 10.1103/RevModPhys.55.403} {\bibfield  {journal}
  {\bibinfo  {journal} {Rev. Mod. Phys.}\ }\textbf {\bibinfo {volume} {55}},\
  \bibinfo {pages} {403} (\bibinfo {year} {1983})}\BibitemShut {NoStop}%
\bibitem [{\citenamefont {Birdsall}\ and\ \citenamefont
  {Langdon}(1991)}]{birdsall}%
  \BibitemOpen
  \bibfield  {author} {\bibinfo {author} {\bibfnamefont {C.}~\bibnamefont
  {Birdsall}}\ and\ \bibinfo {author} {\bibfnamefont {A.}~\bibnamefont
  {Langdon}},\ }\href@noop {} {\emph {\bibinfo {title} {Plasma Physics via
  Computer Simulation}}},\ \bibinfo {edition} {1st}\ ed.\ (\bibinfo
  {publisher} {Taylor and Francis},\ \bibinfo {year} {1991})\BibitemShut
  {NoStop}%
\bibitem [{\citenamefont {Silva}\ \emph {et~al.}(2004)\citenamefont {Silva},
  \citenamefont {Marti}, \citenamefont {Davies}, \citenamefont {Fonseca},
  \citenamefont {Ren}, \citenamefont {Tsung},\ and\ \citenamefont
  {Mori}}]{silva:2004}%
  \BibitemOpen
  \bibfield  {author} {\bibinfo {author} {\bibfnamefont {L.~O.}\ \bibnamefont
  {Silva}}, \bibinfo {author} {\bibfnamefont {M.}~\bibnamefont {Marti}},
  \bibinfo {author} {\bibfnamefont {J.~R.}\ \bibnamefont {Davies}}, \bibinfo
  {author} {\bibfnamefont {R.~A.}\ \bibnamefont {Fonseca}}, \bibinfo {author}
  {\bibfnamefont {C.}~\bibnamefont {Ren}}, \bibinfo {author} {\bibfnamefont
  {F.~S.}\ \bibnamefont {Tsung}}, \ and\ \bibinfo {author} {\bibfnamefont
  {W.~B.}\ \bibnamefont {Mori}},\ }\href {\doibase
  10.1103/PhysRevLett.92.015002} {\bibfield  {journal} {\bibinfo  {journal}
  {Phys. Rev. Lett.}\ }\textbf {\bibinfo {volume} {92}},\ \bibinfo {pages}
  {015002} (\bibinfo {year} {2004})}\BibitemShut {NoStop}%
\bibitem [{\citenamefont {Fiuza}\ \emph {et~al.}(2012)\citenamefont {Fiuza},
  \citenamefont {Fonseca}, \citenamefont {Tonge}, \citenamefont {Mori},\ and\
  \citenamefont {Silva}}]{fiuza:2012}%
  \BibitemOpen
  \bibfield  {author} {\bibinfo {author} {\bibfnamefont {F.}~\bibnamefont
  {Fiuza}}, \bibinfo {author} {\bibfnamefont {R.~A.}\ \bibnamefont {Fonseca}},
  \bibinfo {author} {\bibfnamefont {J.}~\bibnamefont {Tonge}}, \bibinfo
  {author} {\bibfnamefont {W.~B.}\ \bibnamefont {Mori}}, \ and\ \bibinfo
  {author} {\bibfnamefont {L.~O.}\ \bibnamefont {Silva}},\ }\href {\doibase
  10.1103/PhysRevLett.108.235004} {\bibfield  {journal} {\bibinfo  {journal}
  {Phys. Rev. Lett.}\ }\textbf {\bibinfo {volume} {108}},\ \bibinfo {pages}
  {235004} (\bibinfo {year} {2012})}\BibitemShut {NoStop}%
\bibitem [{\citenamefont {Tabak}\ \emph {et~al.}(1994)\citenamefont {Tabak},
  \citenamefont {Hammer}, \citenamefont {Glinsky}, \citenamefont {Kruer},
  \citenamefont {Wilks}, \citenamefont {Woodworth}, \citenamefont {Campbell},
  \citenamefont {Perry},\ and\ \citenamefont {Mason}}]{tabak:1994}%
  \BibitemOpen
  \bibfield  {author} {\bibinfo {author} {\bibfnamefont {M.}~\bibnamefont
  {Tabak}}, \bibinfo {author} {\bibfnamefont {J.}~\bibnamefont {Hammer}},
  \bibinfo {author} {\bibfnamefont {M.~E.}\ \bibnamefont {Glinsky}}, \bibinfo
  {author} {\bibfnamefont {W.~L.}\ \bibnamefont {Kruer}}, \bibinfo {author}
  {\bibfnamefont {S.~C.}\ \bibnamefont {Wilks}}, \bibinfo {author}
  {\bibfnamefont {J.}~\bibnamefont {Woodworth}}, \bibinfo {author}
  {\bibfnamefont {E.~M.}\ \bibnamefont {Campbell}}, \bibinfo {author}
  {\bibfnamefont {M.~D.}\ \bibnamefont {Perry}}, \ and\ \bibinfo {author}
  {\bibfnamefont {R.~J.}\ \bibnamefont {Mason}},\ }\href@noop {} {\bibfield
  {journal} {\bibinfo  {journal} {Physics of Plasmas}\ }\textbf {\bibinfo
  {volume} {1}},\ \bibinfo {pages} {1626} (\bibinfo {year} {1994})}\BibitemShut
  {NoStop}%
\bibitem [{\citenamefont {Chrisman}, \citenamefont {Sentoku},\ and\
  \citenamefont {Kemp}(2008)}]{chrisman:2008}%
  \BibitemOpen
  \bibfield  {author} {\bibinfo {author} {\bibfnamefont {B.}~\bibnamefont
  {Chrisman}}, \bibinfo {author} {\bibfnamefont {Y.}~\bibnamefont {Sentoku}}, \
  and\ \bibinfo {author} {\bibfnamefont {A.}~\bibnamefont {Kemp}},\ }\href@noop
  {} {\bibfield  {journal} {\bibinfo  {journal} {Physics of Plasmas}\ }\textbf
  {\bibinfo {volume} {15}} (\bibinfo {year} {2008})}\BibitemShut {NoStop}%
\bibitem [{\citenamefont {Tonge}\ \emph {et~al.}(2009)\citenamefont {Tonge},
  \citenamefont {May}, \citenamefont {Mori}, \citenamefont {Fiuza},
  \citenamefont {Martins}, \citenamefont {Fonseca},\ and\ \citenamefont
  {Silva}}]{tonge:2009}%
  \BibitemOpen
  \bibfield  {author} {\bibinfo {author} {\bibfnamefont {J.}~\bibnamefont
  {Tonge}}, \bibinfo {author} {\bibfnamefont {J.}~\bibnamefont {May}}, \bibinfo
  {author} {\bibfnamefont {W.~B.}\ \bibnamefont {Mori}}, \bibinfo {author}
  {\bibfnamefont {F.}~\bibnamefont {Fiuza}}, \bibinfo {author} {\bibfnamefont
  {S.~F.}\ \bibnamefont {Martins}}, \bibinfo {author} {\bibfnamefont {R.~A.}\
  \bibnamefont {Fonseca}}, \ and\ \bibinfo {author} {\bibfnamefont {L.~O.}\
  \bibnamefont {Silva}},\ }\href@noop {} {\bibfield  {journal} {\bibinfo
  {journal} {Physical Review E}\ }\textbf {\bibinfo {volume} {16}} (\bibinfo
  {year} {2009})}\BibitemShut {NoStop}%
\bibitem [{\citenamefont {Kemp}, \citenamefont {Cohen},\ and\ \citenamefont
  {Divol}(2010)}]{kemp:2010}%
  \BibitemOpen
  \bibfield  {author} {\bibinfo {author} {\bibfnamefont {A.}~\bibnamefont
  {Kemp}}, \bibinfo {author} {\bibfnamefont {B.}~\bibnamefont {Cohen}}, \ and\
  \bibinfo {author} {\bibfnamefont {L.}~\bibnamefont {Divol}},\ }\href
  {\doibase 10.1063/1.3312825} {\bibfield  {journal} {\bibinfo  {journal}
  {Physics of Plasmas}\ }\textbf {\bibinfo {volume} {17}} (\bibinfo {year}
  {2010}),\ 10.1063/1.3312825}\BibitemShut {NoStop}%
\bibitem [{\citenamefont {May}\ \emph {et~al.}(2011)\citenamefont {May},
  \citenamefont {Tonge}, \citenamefont {Fiuza}, \citenamefont {Fonseca},
  \citenamefont {Silva}, \citenamefont {Ren},\ and\ \citenamefont
  {Mori}}]{may:2011}%
  \BibitemOpen
  \bibfield  {author} {\bibinfo {author} {\bibfnamefont {J.}~\bibnamefont
  {May}}, \bibinfo {author} {\bibfnamefont {J.}~\bibnamefont {Tonge}}, \bibinfo
  {author} {\bibfnamefont {F.}~\bibnamefont {Fiuza}}, \bibinfo {author}
  {\bibfnamefont {R.~A.}\ \bibnamefont {Fonseca}}, \bibinfo {author}
  {\bibfnamefont {L.~O.}\ \bibnamefont {Silva}}, \bibinfo {author}
  {\bibfnamefont {C.}~\bibnamefont {Ren}}, \ and\ \bibinfo {author}
  {\bibfnamefont {W.~B.}\ \bibnamefont {Mori}},\ }\href {\doibase
  10.1103/PhysRevE.84.025401} {\bibfield  {journal} {\bibinfo  {journal} {Phys.
  Rev. E}\ }\textbf {\bibinfo {volume} {84}},\ \bibinfo {pages} {025401}
  (\bibinfo {year} {2011})}\BibitemShut {NoStop}%
\bibitem [{\citenamefont {Spitkovsky}(2008)}]{spitkovsky:2008}%
  \BibitemOpen
  \bibfield  {author} {\bibinfo {author} {\bibfnamefont {A.}~\bibnamefont
  {Spitkovsky}},\ }\href {http://stacks.iop.org/1538-4357/673/i=1/a=L39}
  {\bibfield  {journal} {\bibinfo  {journal} {The Astrophysical Journal
  Letters}\ }\textbf {\bibinfo {volume} {673}},\ \bibinfo {pages} {L39}
  (\bibinfo {year} {2008})}\BibitemShut {NoStop}%
\bibitem [{\citenamefont {Martins}\ \emph {et~al.}(2009)\citenamefont
  {Martins}, \citenamefont {Fonseca}, \citenamefont {Silva},\ and\
  \citenamefont {Mori}}]{martins:2009}%
  \BibitemOpen
  \bibfield  {author} {\bibinfo {author} {\bibfnamefont {S.~F.}\ \bibnamefont
  {Martins}}, \bibinfo {author} {\bibfnamefont {R.~A.}\ \bibnamefont
  {Fonseca}}, \bibinfo {author} {\bibfnamefont {L.~O.}\ \bibnamefont {Silva}},
  \ and\ \bibinfo {author} {\bibfnamefont {W.~B.}\ \bibnamefont {Mori}},\
  }\href {http://stacks.iop.org/1538-4357/695/i=2/a=L189} {\bibfield  {journal}
  {\bibinfo  {journal} {The Astrophysical Journal Letters}\ }\textbf {\bibinfo
  {volume} {695}},\ \bibinfo {pages} {L189} (\bibinfo {year}
  {2009})}\BibitemShut {NoStop}%
\bibitem [{\citenamefont {Sentoku}\ and\ \citenamefont
  {Kemp}(2008)}]{sentoku:2008}%
  \BibitemOpen
  \bibfield  {author} {\bibinfo {author} {\bibfnamefont {Y.}~\bibnamefont
  {Sentoku}}\ and\ \bibinfo {author} {\bibfnamefont {A.~J.}\ \bibnamefont
  {Kemp}},\ }\href@noop {} {\bibfield  {journal} {\bibinfo  {journal} {J. Comp.
  Phys.}\ }\textbf {\bibinfo {volume} {15}},\ \bibinfo {pages} {056309}
  (\bibinfo {year} {2008})}\BibitemShut {NoStop}%
\bibitem [{\citenamefont {Tonge}\ \emph {et~al.}(2010)\citenamefont {Tonge},
  \citenamefont {May}, \citenamefont {Mori}, \citenamefont {Fiuza},\ and\
  \citenamefont {Silva}}]{tonge:2010}%
  \BibitemOpen
  \bibfield  {author} {\bibinfo {author} {\bibfnamefont {J.}~\bibnamefont
  {Tonge}}, \bibinfo {author} {\bibfnamefont {J.}~\bibnamefont {May}}, \bibinfo
  {author} {\bibfnamefont {W.~B.}\ \bibnamefont {Mori}}, \bibinfo {author}
  {\bibfnamefont {F.}~\bibnamefont {Fiuza}}, \ and\ \bibinfo {author}
  {\bibfnamefont {L.~O.}\ \bibnamefont {Silva}},\ }in\ \href@noop {} {\emph
  {\bibinfo {booktitle} {Proceedings of the 40th Annual Anomalous Absorption
  Conference}}}\ (\bibinfo {year} {2010})\BibitemShut {NoStop}%
\bibitem [{\citenamefont {Kemp}\ \emph {et~al.}(2013)\citenamefont {Kemp},
  \citenamefont {Fiuza}, \citenamefont {Debayle}, \citenamefont {Johzaki},
  \citenamefont {Mori}, \citenamefont {Patel}, \citenamefont {Sentoku},\ and\
  \citenamefont {Silva}}]{kemp:2013x}%
  \BibitemOpen
  \bibfield  {author} {\bibinfo {author} {\bibfnamefont {A.~J.}\ \bibnamefont
  {Kemp}}, \bibinfo {author} {\bibfnamefont {F.}~\bibnamefont {Fiuza}},
  \bibinfo {author} {\bibfnamefont {A.}~\bibnamefont {Debayle}}, \bibinfo
  {author} {\bibfnamefont {T.}~\bibnamefont {Johzaki}}, \bibinfo {author}
  {\bibfnamefont {W.~B.}\ \bibnamefont {Mori}}, \bibinfo {author}
  {\bibfnamefont {P.~K.}\ \bibnamefont {Patel}}, \bibinfo {author}
  {\bibfnamefont {Y.}~\bibnamefont {Sentoku}}, \ and\ \bibinfo {author}
  {\bibfnamefont {L.~O.}\ \bibnamefont {Silva}},\ }\href
  {http://arxiv.org/abs/1308.2628} {\bibfield  {journal} {\bibinfo  {journal}
  {arxiv:1308.2628}\ } (\bibinfo {year} {2013})}\BibitemShut {NoStop}%
\bibitem [{\citenamefont {Kemp}, \citenamefont {Sentoku},\ and\ \citenamefont
  {Tabak}(2009)}]{kemp:2009}%
  \BibitemOpen
  \bibfield  {author} {\bibinfo {author} {\bibfnamefont {A.}~\bibnamefont
  {Kemp}}, \bibinfo {author} {\bibfnamefont {Y.}~\bibnamefont {Sentoku}}, \
  and\ \bibinfo {author} {\bibfnamefont {M.}~\bibnamefont {Tabak}},\
  }\href@noop {} {\bibfield  {journal} {\bibinfo  {journal} {Physics of
  Plasmas}\ }\textbf {\bibinfo {volume} {79}},\ \bibinfo {pages} {066406}
  (\bibinfo {year} {2009})}\BibitemShut {NoStop}%
\bibitem [{\citenamefont {Katsouleas}\ \emph {et~al.}(1987)\citenamefont
  {Katsouleas}, \citenamefont {Wilks}, \citenamefont {Chen}, \citenamefont
  {Dawson},\ and\ \citenamefont {Su}}]{katsouleas:1987}%
  \BibitemOpen
  \bibfield  {author} {\bibinfo {author} {\bibfnamefont {T.}~\bibnamefont
  {Katsouleas}}, \bibinfo {author} {\bibfnamefont {S.}~\bibnamefont {Wilks}},
  \bibinfo {author} {\bibfnamefont {P.}~\bibnamefont {Chen}}, \bibinfo {author}
  {\bibfnamefont {J.~M.}\ \bibnamefont {Dawson}}, \ and\ \bibinfo {author}
  {\bibfnamefont {J.~J.}\ \bibnamefont {Su}},\ }\href@noop {} {\bibfield
  {journal} {\bibinfo  {journal} {Particle Accelerators}\ }\textbf {\bibinfo
  {volume} {22}},\ \bibinfo {pages} {81} (\bibinfo {year} {1987})}\BibitemShut
  {NoStop}%
\bibitem [{\citenamefont {Mori}\ \emph {et~al.}(1988)\citenamefont {Mori},
  \citenamefont {Dawson}, \citenamefont {Joshi}, \citenamefont {Katsouleas},
  \citenamefont {Su},\ and\ \citenamefont {Wilks}}]{mori:1988}%
  \BibitemOpen
  \bibfield  {author} {\bibinfo {author} {\bibfnamefont {W.~B.}\ \bibnamefont
  {Mori}}, \bibinfo {author} {\bibfnamefont {J.~M.}\ \bibnamefont {Dawson}},
  \bibinfo {author} {\bibfnamefont {C.}~\bibnamefont {Joshi}}, \bibinfo
  {author} {\bibfnamefont {T.}~\bibnamefont {Katsouleas}}, \bibinfo {author}
  {\bibfnamefont {J.~J.}\ \bibnamefont {Su}}, \ and\ \bibinfo {author}
  {\bibfnamefont {S.}~\bibnamefont {Wilks}},\ }\href {\doibase
  10.1117/12.965097} {\enquote {\bibinfo {title} {The plasma wakefield
  accelerator},}\ } (\bibinfo {year} {1988})\BibitemShut {NoStop}%
\bibitem [{\citenamefont {Abramowitz}\ and\ \citenamefont
  {Stegun}(1970)}]{abramowitz:1970}%
  \BibitemOpen
  \bibfield  {author} {\bibinfo {author} {\bibfnamefont {M.}~\bibnamefont
  {Abramowitz}}\ and\ \bibinfo {author} {\bibfnamefont {I.}~\bibnamefont
  {Stegun}},\ }\href@noop {} {\emph {\bibinfo {title} {Handbook of mathematical
  functions}}}\ (\bibinfo {year} {1970})\BibitemShut {NoStop}%
\bibitem [{\citenamefont {Lu}\ \emph {et~al.}(2005)\citenamefont {Lu},
  \citenamefont {Huang}, \citenamefont {Zhou}, \citenamefont {Mori},\ and\
  \citenamefont {Katsouleas}}]{lu:2005}%
  \BibitemOpen
  \bibfield  {author} {\bibinfo {author} {\bibfnamefont {W.}~\bibnamefont
  {Lu}}, \bibinfo {author} {\bibfnamefont {C.}~\bibnamefont {Huang}}, \bibinfo
  {author} {\bibfnamefont {M.~M.}\ \bibnamefont {Zhou}}, \bibinfo {author}
  {\bibfnamefont {W.~B.}\ \bibnamefont {Mori}}, \ and\ \bibinfo {author}
  {\bibfnamefont {T.}~\bibnamefont {Katsouleas}},\ }\href {\doibase
  http://dx.doi.org/10.1063/1.1905587} {\bibfield  {journal} {\bibinfo
  {journal} {Physics of Plasmas}\ }\textbf {\bibinfo {volume} {12}},\ \bibinfo
  {eid} {063101} (\bibinfo {year} {2005})}\BibitemShut {NoStop}%
\bibitem [{\citenamefont {Fonseca}\ \emph {et~al.}(2002)\citenamefont
  {Fonseca}, \citenamefont {Silva}, \citenamefont {Tsung}, \citenamefont
  {Decyk}, \citenamefont {Lu}, \citenamefont {Ren}, \citenamefont {Mori},
  \citenamefont {Deng}, \citenamefont {Lee}, \citenamefont {Katsouleas},\ and\
  \citenamefont {Adam}}]{fonseca:2002}%
  \BibitemOpen
  \bibfield  {author} {\bibinfo {author} {\bibfnamefont {R.~A.}\ \bibnamefont
  {Fonseca}}, \bibinfo {author} {\bibfnamefont {L.~O.}\ \bibnamefont {Silva}},
  \bibinfo {author} {\bibfnamefont {F.~S.}\ \bibnamefont {Tsung}}, \bibinfo
  {author} {\bibfnamefont {V.~K.}\ \bibnamefont {Decyk}}, \bibinfo {author}
  {\bibfnamefont {W.}~\bibnamefont {Lu}}, \bibinfo {author} {\bibfnamefont
  {C.}~\bibnamefont {Ren}}, \bibinfo {author} {\bibfnamefont {W.~B.}\
  \bibnamefont {Mori}}, \bibinfo {author} {\bibfnamefont {S.}~\bibnamefont
  {Deng}}, \bibinfo {author} {\bibfnamefont {S.}~\bibnamefont {Lee}}, \bibinfo
  {author} {\bibfnamefont {T.}~\bibnamefont {Katsouleas}}, \ and\ \bibinfo
  {author} {\bibfnamefont {J.~C.}\ \bibnamefont {Adam}},\ }\href@noop {}
  {\bibfield  {journal} {\bibinfo  {journal} {Lecture Notes in Computer
  Science}\ }\textbf {\bibinfo {volume} {2331}},\ \bibinfo {pages} {342}
  (\bibinfo {year} {2002})}\BibitemShut {NoStop}%
\bibitem [{\citenamefont {Fonseca}\ \emph {et~al.}(2008)\citenamefont
  {Fonseca}, \citenamefont {Martins}, \citenamefont {Silva}, \citenamefont
  {Tonge}, \citenamefont {Tsung},\ and\ \citenamefont {Mori}}]{fonseca:2008}%
  \BibitemOpen
  \bibfield  {author} {\bibinfo {author} {\bibfnamefont {R.~A.}\ \bibnamefont
  {Fonseca}}, \bibinfo {author} {\bibfnamefont {S.~F.}\ \bibnamefont
  {Martins}}, \bibinfo {author} {\bibfnamefont {L.~O.}\ \bibnamefont {Silva}},
  \bibinfo {author} {\bibfnamefont {J.~W.}\ \bibnamefont {Tonge}}, \bibinfo
  {author} {\bibfnamefont {F.~S.}\ \bibnamefont {Tsung}}, \ and\ \bibinfo
  {author} {\bibfnamefont {W.~B.}\ \bibnamefont {Mori}},\ }\href
  {http://stacks.iop.org/0741-3335/50/124034} {\bibfield  {journal} {\bibinfo
  {journal} {Plasma Physics and Controlled Fusion}\ }\textbf {\bibinfo {volume}
  {50}},\ \bibinfo {pages} {124034 (9pp)} (\bibinfo {year} {2008})}\BibitemShut
  {NoStop}%
\bibitem [{\citenamefont {Fonseca}\ \emph {et~al.}(2013)\citenamefont
  {Fonseca}, \citenamefont {Vieira}, \citenamefont {Fiuza}, \citenamefont
  {Davidson}, \citenamefont {Tsung}, \citenamefont {Mori},\ and\ \citenamefont
  {Silva}}]{fonseca:2013}%
  \BibitemOpen
  \bibfield  {author} {\bibinfo {author} {\bibfnamefont {R.~A.}\ \bibnamefont
  {Fonseca}}, \bibinfo {author} {\bibfnamefont {J.}~\bibnamefont {Vieira}},
  \bibinfo {author} {\bibfnamefont {F.}~\bibnamefont {Fiuza}}, \bibinfo
  {author} {\bibfnamefont {A.}~\bibnamefont {Davidson}}, \bibinfo {author}
  {\bibfnamefont {F.~S.}\ \bibnamefont {Tsung}}, \bibinfo {author}
  {\bibfnamefont {W.~B.}\ \bibnamefont {Mori}}, \ and\ \bibinfo {author}
  {\bibfnamefont {L.~O.}\ \bibnamefont {Silva}},\ }\href
  {http://stacks.iop.org/0741-3335/55/i=12/a=124011} {\bibfield  {journal}
  {\bibinfo  {journal} {Plasma Physics and Controlled Fusion}\ }\textbf
  {\bibinfo {volume} {55}},\ \bibinfo {pages} {124011} (\bibinfo {year}
  {2013})}\BibitemShut {NoStop}%
\bibitem [{\citenamefont {Tonge}(2002)}]{parsec}%
  \BibitemOpen
  \bibfield  {author} {\bibinfo {author} {\bibfnamefont {J.}~\bibnamefont
  {Tonge}},\ }\href@noop {} {Ph.D. thesis},\ \bibinfo  {school} {UCLA}
  (\bibinfo {year} {2002})\BibitemShut {NoStop}%
\bibitem [{\citenamefont {Decyk}(1987)}]{decyk:1987}%
  \BibitemOpen
  \bibfield  {author} {\bibinfo {author} {\bibfnamefont {V.}~\bibnamefont
  {Decyk}},\ }in\ \href@noop {} {\emph {\bibinfo {booktitle} {Int. Conf. on
  Plasma Physics (Kiev, USSR)}}}\ (\bibinfo  {publisher} {Singapure: World
  Scientific},\ \bibinfo {year} {1987})\BibitemShut {NoStop}%
\bibitem [{\citenamefont {Grismayer}\ \emph {et~al.}(2011)\citenamefont
  {Grismayer}, \citenamefont {Fahlen}, \citenamefont {Decyk},\ and\
  \citenamefont {Mori}}]{grismayer:2011}%
  \BibitemOpen
  \bibfield  {author} {\bibinfo {author} {\bibfnamefont {T.}~\bibnamefont
  {Grismayer}}, \bibinfo {author} {\bibfnamefont {J.~E.}\ \bibnamefont
  {Fahlen}}, \bibinfo {author} {\bibfnamefont {V.~K.}\ \bibnamefont {Decyk}}, \
  and\ \bibinfo {author} {\bibfnamefont {W.~B.}\ \bibnamefont {Mori}},\ }\href
  {http://stacks.iop.org/0741-3335/53/i=7/a=074011} {\bibfield  {journal}
  {\bibinfo  {journal} {Plasma Physics and Controlled Fusion}\ }\textbf
  {\bibinfo {volume} {53}},\ \bibinfo {pages} {074011} (\bibinfo {year}
  {2011})}\BibitemShut {NoStop}%
\bibitem [{\citenamefont {Takizuka}\ and\ \citenamefont
  {Abe}(1977)}]{takizuka:1977}%
  \BibitemOpen
  \bibfield  {author} {\bibinfo {author} {\bibfnamefont {T.}~\bibnamefont
  {Takizuka}}\ and\ \bibinfo {author} {\bibfnamefont {H.}~\bibnamefont {Abe}},\
  }\href@noop {} {\bibfield  {journal} {\bibinfo  {journal} {J. Comp. Phy.}\
  }\textbf {\bibinfo {volume} {25}},\ \bibinfo {pages} {205} (\bibinfo {year}
  {1977})}\BibitemShut {NoStop}%
\bibitem [{\citenamefont {Nambu}\ and\ \citenamefont
  {Yonemura}(1998)}]{nambu:1998}%
  \BibitemOpen
  \bibfield  {author} {\bibinfo {author} {\bibfnamefont {K.}~\bibnamefont
  {Nambu}}\ and\ \bibinfo {author} {\bibfnamefont {S.}~\bibnamefont
  {Yonemura}},\ }\href@noop {} {\bibfield  {journal} {\bibinfo  {journal} {J.
  Comp. Phy.}\ }\textbf {\bibinfo {volume} {145}},\ \bibinfo {pages} {639}
  (\bibinfo {year} {1998})}\BibitemShut {NoStop}%
\bibitem [{\citenamefont {Peano}\ \emph {et~al.}(2009)\citenamefont {Peano},
  \citenamefont {Marti}, \citenamefont {Silva},\ and\ \citenamefont
  {Coppa}}]{peano:2009}%
  \BibitemOpen
  \bibfield  {author} {\bibinfo {author} {\bibfnamefont {F.}~\bibnamefont
  {Peano}}, \bibinfo {author} {\bibfnamefont {M.}~\bibnamefont {Marti}},
  \bibinfo {author} {\bibfnamefont {L.~O.}\ \bibnamefont {Silva}}, \ and\
  \bibinfo {author} {\bibfnamefont {G.}~\bibnamefont {Coppa}},\ }\href
  {\doibase 10.1103/PhysRevE.79.025701} {\bibfield  {journal} {\bibinfo
  {journal} {Phys. Rev. E}\ }\textbf {\bibinfo {volume} {79}},\ \bibinfo
  {pages} {025701} (\bibinfo {year} {2009})}\BibitemShut {NoStop}%
\bibitem [{\citenamefont {Kato}(2013)}]{kato:2013}%
  \BibitemOpen
  \bibfield  {author} {\bibinfo {author} {\bibfnamefont {T.~N.}\ \bibnamefont
  {Kato}},\ }\href {http://arxiv.org/abs/1312.5507} {\bibfield  {journal}
  {\bibinfo  {journal} {arxiv:1312.5507}\ } (\bibinfo {year}
  {2013})}\BibitemShut {NoStop}%
\end{thebibliography}%

\newpage


\end{document}